\documentclass[12pt,preprint]{aastex}
\usepackage{emulateapj5,apjfonts}
\usepackage[obeyspaces]{url}
\usepackage{graphics}
\usepackage{amssymb}
\usepackage{bbm}
\usepackage{amsmath,textcomp,array}
\usepackage{onecolfloat}
\usepackage{bm}
\usepackage[flushleft]{threeparttable}
\usepackage{xcolor}

\lefthead{Sultan et al.}
\righthead{The Last Journey. II.}

\begin{document}

\def\head{

\title{The Last Journey. II. SMACC -- Subhalo Mass-loss Analysis using Core Catalogs}
\author{Imran Sultan\altaffilmark{1,2}, Nicholas Frontiere\altaffilmark{2,3}, Salman Habib\altaffilmark{2,3}, Katrin Heitmann\altaffilmark{2}, Eve Kovacs\altaffilmark{2}, Patricia Larsen\altaffilmark{2,3}, Esteban Rangel\altaffilmark{4}}

\affil{$^1$ Department of Physics \& Astronomy and Center for Interdisciplinary Exploration and Research in Astrophysics, Northwestern University, 1800 Sherman Ave, Evanston, IL 60201}

\affil{$^2$ High Energy Physics Division,  Argonne National Laboratory, Lemont, IL 60439}

\affil{$^3$ Computational Science Division,  Argonne National Laboratory, Lemont, IL 60439}

\affil{$^4$ Argonne Leadership Computing Facility,  Argonne National Laboratory, Lemont, IL 60439}

\date{today}

\begin{abstract}

This paper introduces SMACC -- Subhalo Mass-loss Analysis using Core Catalogs. SMACC adds a mass model to substructure merger trees based on halo ``core tracking.'' Our approach avoids the need for running expensive subhalo finding algorithms and instead uses subhalo mass-loss modeling to assign masses to halo cores. We present details of the SMACC methodology and demonstrate its excellent performance in describing halo substructure and its evolution. Validation of the approach is carried out using cosmological simulations at significantly different resolutions. We apply SMACC to the 1.24 trillion-particle Last Journey simulation and construct core catalogs with the additional mass information. These catalogs can be readily used as input to semi-analytic models or subhalo abundance matching approaches to determine approximate galaxy distributions, as well as for in-depth studies of small-scale structure evolution.

\end{abstract}

\keywords{methods: N-body ---
          cosmology: large-scale structure of the universe}}


\twocolumn[\head]

\section{Introduction}

Cosmological simulations have provided, and continue to provide, otherwise wholly inaccessible insights into the highly nonlinear process of large-scale structure formation in the Universe (for reviews, see, e.g., \citealt{Bertschinger98, Dolag2008, Frenk2012}). The volume and resolution achieved by these simulations has improved tremendously over the last few decades following the trajectory for computational power set by Moore's Law. This has allowed for the emergence of an ever more detailed picture of the complex structures formed by the action of gravity in an expanding universe. 

In the fiducial $\Lambda$CDM cosmology, structure formation is hierarchical, with small halos -- localized clumps of matter -- forming first and then merging into larger halos, which then in turn merge into even larger ones, a process captured by the notion of merger trees~\citep{laceycole93}. Such a dynamical evolution naturally introduces the concept of halo substructure as smaller halos fall into more massive objects and live out a (potentially perilous) existence as subhalos, subject to tidal disruption and mergers. Further substructure exists within subhalos as well, complicating the analysis of high-resolution simulations.

If one is interested only in the density field associated with halos, there is no particular reason to be concerned about identifying which region is a piece of the main halo and which corresponds to substructure, as long as the density field is sufficiently resolved. More information, however, is required for further detailed analyses:  Investigations of cosmological probes based on large-scale structure require modeling galaxy locations, as well as other properties of galaxies. Therefore, in the context of a gravity-only N-body simulation, it becomes necessary to model galaxy formation and evolution by using detailed spatio-temporal knowledge of the properties of halos, halo substructure, and the halo environment. Such an approach, referred to as the 
 ``galaxy-halo connection'', can be studied via various combinations of N-body simulation results, empirical modeling (e.g., Halo Occupation Distribution (HOD), subhalo abundance matching (SHAM)) and semi-analytic modeling (SAM) techniques, including modeling of gas dynamics and astrophysical feedback effects (for reviews, see \citealt{mo11,somerville15,wechsler18}).

In this picture, the association of galaxies with subhalos is nontrivial because of the complexity of the ever-evolving picture of ``structures within structures.'' As one example, observations of group and cluster-scale halos have shown that the radial distribution of galaxies does not follow that of simulated subhalos~\citep{lin2004,budzynski12}. For small-scale clustering measurements, related challenges have been discussed in~\cite{campbell18}. The one-to-one association of galaxies with their host subhalos can be lost due to physical mechanisms such as tidal stripping, and in simulations, due to problems with mass resolution and subhalo finding (especially near halo centers). Galaxies that are no longer associated with their dark matter subhalo are termed ``orphan'' galaxies; dealing properly with orphans is an important aspect of semi-analytic galaxy modeling.

The short discussion above points to the importance of robustly tracking substructure in cosmological simulations. Identification of substructure is a challenging computational task \citep{muldrew11}, however, and made more difficult by the lack of a rigorous definition of halos and their substructure and by the lack of direct observational analogs. Indeed, there are many approaches to subhalo identification and tracking in simulations, all of which have their own features and strengths and weaknesses (for a detailed comparison project, see \citealt{onions}). In addition, the requirements on the force and mass resolution for the simulations to ensure that subhalos are captured reliably poses a major challenge (see, e.g., \citealt{guo2014, vdB17,vdBO18} for recent discussions). 

In order to address some important aspects of substructure identification and the related problem of galaxy-substructure association, we have recently introduced the concept of ``core tracking'' in N-body simulations~\citep{rangel17,korytov,LJ1}. The intellectual ancestry of this idea goes back to \cite{kaiser84} and \cite{white87}, focusing on following substructure peaks as potential galaxy locations. In our present-day context, a core is defined by identifying a set of particles closest to the center of a halo, the process beginning when a halo reaches some initial mass threshold. Once found, all core particles are tracked for the remainder of the simulation. By combining the core particle information with detailed halo merger trees, one can build core catalogs that follow the cores from their birth to the final step in the simulation. Two advantages of core tracking that address the aforementioned difficulties of subhalo finding are: 1) cores are never lost because they are tracked forever and 2) core tracking requires only list construction and comparison, making the approach computationally inexpensive compared to subhalo-based methods.  

The core tracking approach has already been successfully applied to SHAM models in \cite{korytov}, which demonstrated the capability to model dynamical processes such as galaxy mergers and disruption, circumventing the problem of missing orphan galaxies inherent with subhalo finding.
However, core tracking is by no means a complete description of substructure: Once a halo has merged into another halo and become an internal structure, its identity becomes uncertain and its properties are not easily measurable -- tracking the original halo core is, at first sight, not obviously of direct help in reconstructing, for example, the evolution of the associated subhalo mass. Fortunately, previous work has shown that the evolution of subhalo masses in simulations appears, on average, to follow simple analytic forms that characterise the mass loss as the subhalos orbit around a host object \citep{vdB2005,giocoli08, JvdB2016a, vdB16, han18, hiroshima18}. While the reality is complicated by other dynamics, \cite{vdB2005} have been able to reproduce the subhalo mass function (SHMF) from initial substructure masses originally developed in the context of Extended Press-Schechter (EPS) trees. 
Building upon these studies we can determine a simple mass-loss model for predicting the evolution of the subhalo mass associated with merged cores.

In this paper, we introduce SMACC, Subhalo Mass-loss Analysis using Core Catalogs, which allows us to build core catalogs that not only contain the full merger history for each core
but also model the evolution of its associated mass beyond the time it has merged with a more massive object. The resulting `core masses' can be used as inputs to SAMs to create synthetic galaxy catalogs. SAMs traditionally use subhalos to track and determine galaxy properties and require specific subhalo information that is supplied from subhalo merger trees. 
The information inputs needed are the positions, velocities, masses, and evolution history of the subhalos, as well as a range of measurements for the infalling halo and the host halo that contains the subhalo. In order to use cores as a substitute for subhalos, we need to provide the analogs of these quantities for cores. Positions and velocities of the cores and properties of the infall and host halos are already described by the core merger trees by construction. SMACC provides the additional mass information needed to enable galaxy modeling using SAMs.  

This paper is the second in a series of papers focused on the Last Journey simulation. The simulation, described in the first paper~\citep{LJ1}, is a large, gravity-only run with the Hardware/Hybrid Accelerated Cosmology Code (HACC) \citep{habib14} designed to enable the creation of detailed synthetic sky catalogs at different wavelengths. In the current paper, we focus on expanding the Last Journey core catalogs to generate a suitable set of inputs for SAM methods using SMACC. We investigate the robustness of subhalo mass modeling via SMACC as a function of mass resolution, the host halo concentration, individual subhalo-to-core matching, and also to variations in cosmological parameters, finding that the underlying mass-loss model provides a very satisfactory description in all cases studied.

This paper is organized as follows. We begin in Section~\ref{sec:simulations} with a brief overview of the simulations used in this paper; we additionally provide a short description of our subhalo finder and show that it yields results in good agreement with current literature. In Section~\ref{sec:coreconcept}, we review the definition of cores and how merger trees are used to enable core tracking. We demonstrate in Section~\ref{sec:massmodeling} that a simple two-parameter mass-loss model for cores can be used to accurately predict the SHMF by comparing to results obtained with a subhalo finder. In Section~\ref{sec:validation}, we discuss a range of measurements obtained with SMACC to verify the robustness of our approach. We present the results for our extreme-scale simulation, the Last Journey in Section~\ref{sec:results}, and finally provide a summary of the results and future outlook in Section~\ref{sec:summary}. 

\section{Simulations and Subhalos}
\label{sec:simulations}

\begin{table*}
{
\centering
\setlength\tabcolsep{2.0pt}
\caption{Simulations used in this paper}
\begin{tabular}{c|c|c|c|c|c|c|c}
     Name & Simulation Volume & Particle & Particle mass  & Core host halo mass& Core size & Host halo mass & Subhalo mass\\
      & [($h^{-1}$Mpc)$^3$] & count & [$h^{-1}$M$_\odot$]  & [$h^{-1}$M$_\odot$]& & [$h^{-1}$M$_\odot$] & [$h^{-1}$M$_\odot$]  \\
     \hline\hline
     Last Journey & 3400$^3$ & 10,752$^3$ & $\sim 2.7\cdot 10^9$ & > $2.2\cdot 10^{11}$ (80) & 50 & N/A & N/A\\
     Last Journey-SV & 250$^3$ & 1024$^3$ & $\sim 1.3\cdot10^9$ & > $2.5\cdot 10^{10}$ (20) & 20 & > $8.8\cdot 10^{10}$ (70) & > $2.5\cdot 10^{10}$ (20)\\
     Last Journey-HM & 250$^3$ & 3072$^3$ & $\sim 4.6\cdot 10^7$ & > $3.7\cdot 10^{9}$ (80) &  50 & > $9.3\cdot 10^{10}$ (2,000) & > $9.3\cdot 10^{8}$ (20)\\
     AlphaQ & 256$^3$ & 1024$^3$ & $\sim 1.2\cdot 10^9$ & > $2.4\cdot 10^{10}$ (20)&  20 & > $8.4\cdot 10^{10}$ (70) & > $2.4\cdot 10^{10}$ (20)
\end{tabular}\label{tab:sims}
\begin{tablenotes}
          \item 
          \small Summary of the specifications of the four simulations used in this paper. The first four columns specify the simulations and their properties (simulation identifier, volume, particle count, and particle mass). The next four columns list the range of core host halo masses (the minimum halo mass above which a core is associated with a halo), the core size (number of core particles), the range of host halo masses (halos hosting at least one subhalo), and the range of subhalo masses. The numbers in parentheses for the reported mass thresholds are the corresponding number of simulation particles; the host halo mass is an FOF mass with $b=0.168$. The force resolution for the three lower mass resolution simulations is $\sim$2-3$h^{-1}$kpc and for the HM simulation $\sim$0.8$h^{-1}$kpc. The three `Last Journey' simulations are based on Planck-18 best-fit parameters, while the AlphaQ simulation is based on WMAP-7 best-fit parameters. (For more details, see text.)
        \end{tablenotes}
}
\end{table*}

\begin{figure*}[t]
\hspace{0.13cm}\centerline{\includegraphics[width=6.50in]{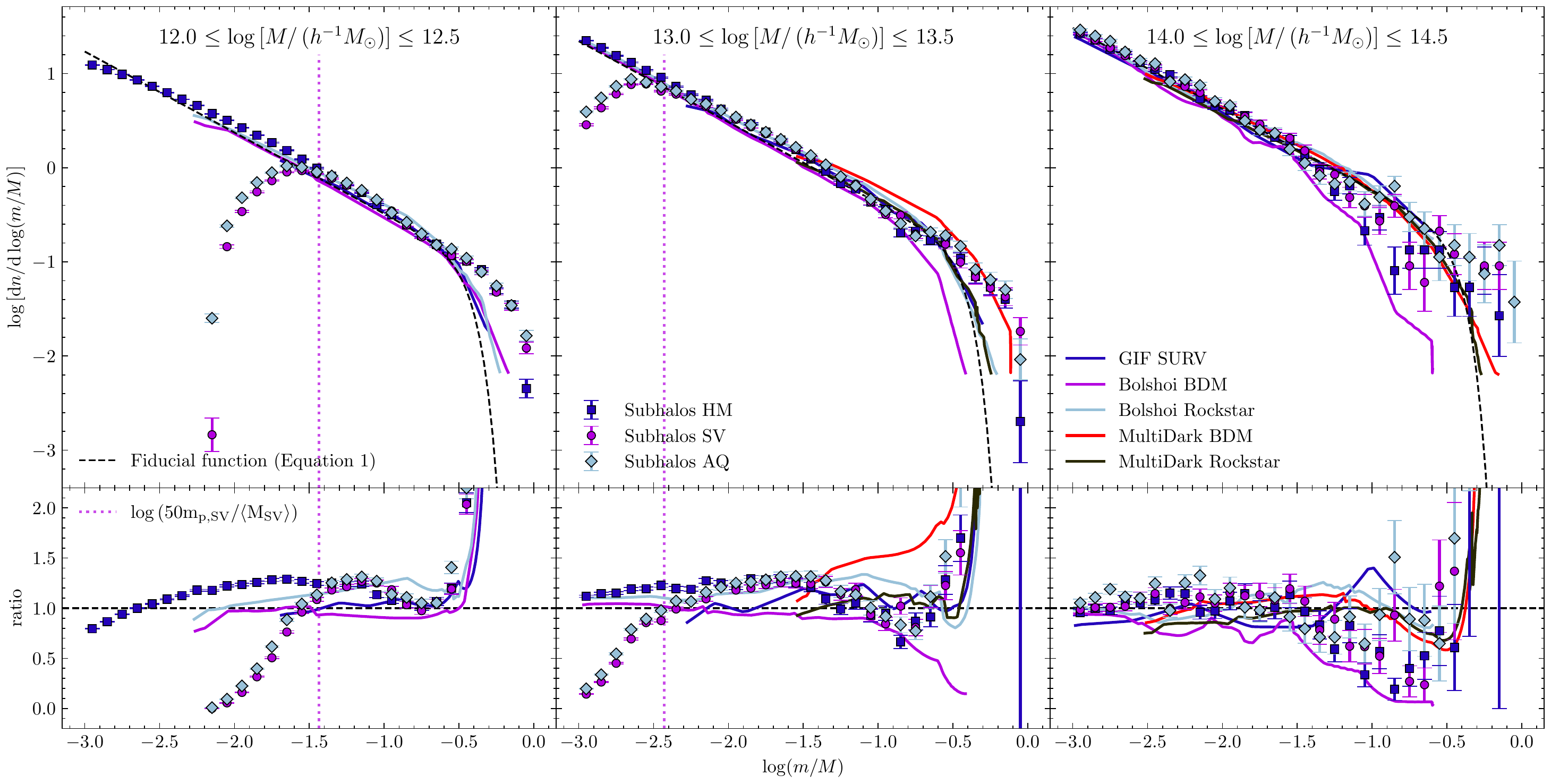}}
\caption{\label{fig:subfinder_comparison} Comparison of our subhalo finder results to those from other implementations. Upper panels: The average subhalo mass function at $z=0$ for three host halo mass bins in the HM (squares), SV (circles) and AlphaQ simulations (diamonds) where subhalos with $m>20$ particles are shown (see text for the number of host halos in each bin). Results from a number of other simulations using different subhalo finders (solid lines) are shown as reported in~\cite{vdB16}. Following \cite{vdB16}, we use Equation \ref{eq:fittingfn} with $\alpha=0.86$ and $A_M=0.045$, 0.058, and 0.080 for the $\log \left[ M / \left(h^{{-1}}\mathrm{M_\odot} \right) \right]\in[12.0,12.5]$, $[13.0,13.5]$, and $[14.0,14.5]$ host bins, respectively, as fiducial fitting functions to compare the results of each bin. The fiducial fitting functions are indicated with dashed lines. The vertical dotted line in the left and middle panels approximates the 50 particle mass threshold for SV, following~\cite{vdB16}.
Lower panels: Ratio of the subhalo mass functions to the fiducial fitting functions (dashed lines) shown in the upper panels. Note, that while we use the FOF finder to identify the host halos, for this plot we have used the corresponding virial mass measurement for determining $M$ to enable a meaningful comparison with the other results.}
\end{figure*} 

In this section, we summarize the simulations utilized in our analysis. The ultimate goal is to enable efficient substructure modeling for very large simulations, in particular the Last Journey simulation presented in~\cite{LJ1}. For development and validation, however, smaller simulations are often very useful since they allow for faster data processing and testing.  In addition, the analysis overhead does not require very large computing allocations or memory footprints. Consequently, we used a suite of smaller scale simulations, described below, to carry out a set of modeling and convergence studies. All of these simulations were performed with the N-body code HACC, described in~\cite{habib14}. HACC has been designed to run at scale on a range of architectures; the simulations used in this paper have been run on IBM BG/Q and GPU-accelerated systems. We will also briefly describe the subhalo finder employed for comparisons to our new core-tracking methodology, and show how well it matches results from other subhalo finders commonly used in the literature. Subhalo finding is carried out for our smaller simulations at a selected number of redshifts. 

The three main simulations used in the paper are based on a Planck-18 cosmology~\citep{planck18} with $\theta=\{\Omega_{\rm cdm}=0.26067, \omega_b=0.02242, h=0.6766, \sigma_8=0.8102, n_s=0.9665,w=-1\}$, assuming a flat universe and massless neutrinos. The final target simulation is the Last Journey run, which evolved 10,752$^3$ particles in a (3400$h^{-1}$Mpc)$^3$ volume. The mass resolution for this simulation is $m_p=2.7\cdot 10^9 h^{-1}$M$_\odot$. 

Two smaller simulations were utilized for validation experiments. For the first run, we simply reduced the volume but kept the mass resolution similar to that in the Last Journey simulation. We chose a (250$h^{-1}$Mpc)$^3$ volume and 1024$^3$ simulation particles, leading to a mass resolution of $m_p=1.252\cdot 10^9 h^{-1}$M$_\odot$. The resulting smaller dataset is much easier to handle and lends itself to a wide range of tests. In this case, to facilitate resolution measurements, we also reduce (to 20) the number of particles required to begin tracking cores (See Table~\ref{tab:sims}, discussed later). 
We will refer to this simulation in the following as Last Journey-SV, or simply SV, for ``small volume.'' For the second run, we generated a simulation with much higher mass resolution, using the same volume of (250$h^{-1}$Mpc)$^3$ but increasing the number of particles to $3072^3$, leading to a mass resolution of $m_p=4.637\cdot 10^7 h^{-1}$M$_\odot$. The main purpose of this simulation is to enable detailed mass resolution and convergence studies for halo substructure. We will refer to this simulation in the following as Last Journey-HM, or simply HM, for high mass resolution. 

Finally, to provide a variation in cosmology, we employ an additional simulation based on WMAP-7 results~\citep{wmap7} in a (256$h^{-1}$Mpc)$^3$ volume, the AlphaQ simulation. The cosmological parameters used for the AlphaQ run are: $\theta=\{\Omega_{\rm cdm}=0.220, \omega_b=0.02258, h=0.71, \sigma_8=0.8, n_s=0.963, w=-1\}$. As for the Last Journey simulations, we assume flatness and massless neutrinos. The AlphaQ simulation is very similar to the Last Journey-SV simulation with regard to volume and mass resolution. The simulation was used in the past to develop a workflow to generate the synthetic galaxy catalog as part of the cosmoDC2 effort described in~\cite{cosmoDC2}.

We summarize the simulation suite used in the paper in Table~\ref{tab:sims}. In addition to the simulation specifications we also provide information about minimum host halo and substructure masses for all simulations. For each simulation we find halos at 101 timesteps between $z=10$ and $z=0$, evenly spaced in $\log a$ intervals (whenever we use $\log$ it stands for $\log_{10})$. For each snapshot, we store a halo catalog with detailed information covering a range of halo properties, including positions, velocities, shapes, and masses. A detailed description of the analysis performed for the Last Journey simulation is given in~\cite{LJ1}. 

\subsection{Subhalo Finder Description and Validation}
\label{sec:SHF}

In order to develop our mass modeling approach on cores and optimize the associated modeling parameters, we use results obtained from an independent subhalo finder run on the same simulation. We make use of the subhalo finder operating within the HACC analysis framework, CosmoTools, as described in~\cite{li16} where it was used to identify substructures for detailed simulation-based strong lensing studies. This subhalo finding approach is based on local density estimation (similar to e.g., {\sc Subfind} by ~\citealt{subfind}), but is then augmented by velocity information, helping to better distinguish whether a particle belongs to a subhalo or is a member of the host halo. 

We first identify all halos in the simulation with a fast, parallel friends-of-friends (FOF) finder with a linking length of $b=0.168$ (smaller than the often used fiducial value of $b=0.2$ to reduce overlinking artifacts). The FOF halo finding approach, based on an approximate isodensity boundary, has a long history in cosmology, and was first introduced by~\cite{davis85}. It is algorithmically well defined and useful down to small particle counts. We then organize the particles that belong to the FOF host halo into a Barnes-Hut tree structure~\citep{barneshut} to enable efficient access. The local density is then estimated for each particle using a smoothed particle hydrodynamics (SPH) kernel. Particles are assigned to subhalo candidates and we build two initial membership lists, one for the main halo and one for the subhalo candidates. For each preliminary subhalo we then determine if all assigned subhalo particles are bound to the substructure. For this step we calculate the escape velocity for each particle with regard to the subhalo, using its total energy. If the particle does not belong to the subhalo, we assign it to the main halo and a new energy calculation is carried out and the remaining particles are analyzed. This process continues until only bound particles remain in each subhalo candidate. Finally, we use a mass threshold to determine if the resulting subhalo candidate is still viable. Subhalos with too few particles are discarded. For each subhalo we record its properties in a subhalo property file (positions, velocities, velocity dispersions and masses as well as subhalo tags and tags to identify its host halo). We also separately save the particles belonging to the subhalos, including the relevant tag identifiers.

For comparison with other subhalo finders, we use the Last Journey-SV and HM simulations as well as the AlphaQ simulation at $z=0$. We measure the subhalo mass function for all three simulations in three mass bins and compare the results to previous findings. In particular, we take advantage of the useful compilation of SHMF measurements presented in~\cite{vdB16}. For the host halos, we impose a mass cut of just under $10^{11} h^{-1}$M$_\odot$ (see Table~\ref{tab:sims} for the exact limits). This mass cut is chosen to capture all halos that are large enough to host at least one subhalo. For the comparison, we also measure the $M_{200c}$ host halo mass and convert it to a virial mass estimate using halo concentration measurements. We identify subhalos with over 20 particles. This last requirement leads to a minimum subhalo mass of over $\sim 2.5\cdot 10^{10}h^{-1}$M$_\odot$ and $\sim 9.3\cdot 10^{8}h^{-1}$M$_\odot$ for the SV and HM simulations, respectively and $\sim 2.4\cdot 10^{10}h^{-1}$M$_\odot$ for the AlphaQ simulation.
The HM simulation allows us to investigate the convergence of the SHMF down to very low masses. We will use the subhalo finder results for $z=1$ and $z=0$ for the SV and HM simulations (with FOF host mass measurements using $b=0.168$) throughout the paper to tune and verify our mass modeling approach for cores. 

The upper panels of Figure~\ref{fig:subfinder_comparison} compare our measurements with subhalo mass functions delivered by different subhalo finders in three mass bins, while the lower panels show the ratios with respect to the fiducial SHMF given in Equation~6 of~\cite{vdB16}: 
\begin{equation}\label{eq:fittingfn}
\frac{\mathrm{d}N}{\mathrm{d}\log(m/M)}=A_M\left(\frac{m}{M}\right)^{-\alpha}\exp\left[-50(m/M)^4\right],
\end{equation}
with $A_M=0.045$, 0.058, and 0.080 for the $\log \left[ M / \left(h^{{-1}}\mathrm{M_\odot} \right) \right]\in[12.0,12.5]$, $[13.0,13.5]$, and $[14.0,14.5]$ host halo bins, respectively, and $\alpha=0.86$ as in their Figure 5.
We indicate the fiducial functions in the upper panels with a dashed line. The symbols with error bars are from our simulations (for the four largest FOF halos from the HM simulation shown in the right panel in the figure, we use an approximate unbinding algorithm to speed up the calculation). The left, middle, and right panels of the figure show bins containing (42183, 41128, 36524), (4965, 4929, 4354), and (373, 331, 267) host halos for the (HM, SV, AlphaQ) simulations, respectively. The solid lines in Figure~\ref{fig:subfinder_comparison} are from~\cite{vdB16}.  We note the generally good agreement of our SHMF with those measured with commonly used subhalo finders. The results are consistent with those reported in the major subhalo finder comparison carried out in \cite{onions}, which found overall agreement for the SHMF at the 20\% level; \cite{vdB16} confirmed these findings as well. 

There are two small differences between the measurement carried out in \cite{vdB16} and our analysis. First, \cite{vdB16} use the virial mass for the host halo mass measurement, which is close to $M_{100c}$ and leads to higher host halo estimates compared to the FOF, $b=0.168$ host halo mass used here. In order to compensate for this difference, we use a virial mass estimate for the host halo mass when presenting our results. We note that since $M_{\rm vir}$ halos are larger than FOF, $b=0.168$ halos, subhalos on the outskirts might be undercounted. Second, the cosmology used in our paper is different. The Bolshoi and MultiDark simulations are based on WMAP results. The Planck-18 parameters used in our paper have slightly higher values for $\sigma_8$ and $\Omega_{\rm cdm}$ and lead to more clustering. However, the comparison of the SHMF from the SV and the AlphaQ simulations (Planck-18 and WMAP-7 cosmologies, respectively) in Figure~\ref{fig:subfinder_comparison} shows that varying $\Lambda$CDM parameters, while staying close to the current best-fit cosmology, has a negligible effect. 

Besides establishing the overall agreement with other subhalo finders, the comparison between the SV and HM results is important for convergence studies. The excellent agreement of the SHMF in the (higher subhalo mass) overlap region shows that the lower mass resolution faithfully captures the subhalos of interest. In the lower range of subhalo masses, the limited mass resolution of the SV simulation results in subhalo loss and the SHMF rolls off. (We note that this resolution limit is well below the subhalo mass cut-off that we are planning to use for further studies with the main Last Journey simulation.) 

\section{Core Tracking}
\label{sec:coreconcept}

We have recently introduced the concept of core tracking in cosmological simulations as an alternative to the construction of subhalo merger trees. A detailed description of the core catalog implementation is reported in ~\cite{rangel17} and ~\cite{LJ1}. The first scientific application of core tracking, in addition to comprehensive validation results, is discussed in~\cite{korytov}. In this section, we provide a brief summary of the underlying concepts and construction techniques for the core catalogs.  

\subsection{Definition of Cores}

The definition of a core is straightforward: For each halo above a certain mass (the core host halo mass threshold), the gravitational potential center is found and the set of particles closest to the center is identified as the halo core. The number of particles that define a core is an input parameter to the analysis set-up and is chosen with regard to the size of the substructure that we aim to capture and the mass resolution of the simulation. The values for the core host halo mass threshold and the core size for our simulations are specified in Table~\ref{tab:sims}. For each (of the 101) analysis time snapshots, all core particles identified are stored, including their halo tag, to enable the connection to the halo catalogs at a later stage.

In addition to the core particle files per snapshot, we also keep track of the evolution of the core particles over time. Once a particle has been identified as a core particle, we follow its path for the remainder of the simulation. In totality, at each analysis step, we record the new positions and velocities of core particles that have been found previously and add newly found core particles as we go along. In this way, we build up an accumulated core particle file over time containing information about each core particle's evolution from its first identification until the final time step. Storing the complete core history is necessary to generate core catalogs (as described in Section \ref{subsec:corecat}), which track the complex formation history of identified substructure.  

\subsection{Merger Trees}
\label{sec:mergertrees}

In order to extract the temporal evolution of the substructure traced by core particles, halo merger trees are required. Such trees track the hierarchical formation history of each halo over time, recording merging events and mass accretion. The merger trees are constructed by processing halo catalogs at adjacent analysis snapshots. We compute the overlap of halo particles in each snapshot pair to connect older {\it progenitor} halos with younger {\it descendant} halos at the later timestep. Processing all snapshots results in a complete merger tree catalog that can extract the formation history of any halo of interest.  

\begin{figure}[t]
\hspace{0.13cm}\centerline{\includegraphics[width=3.6in]{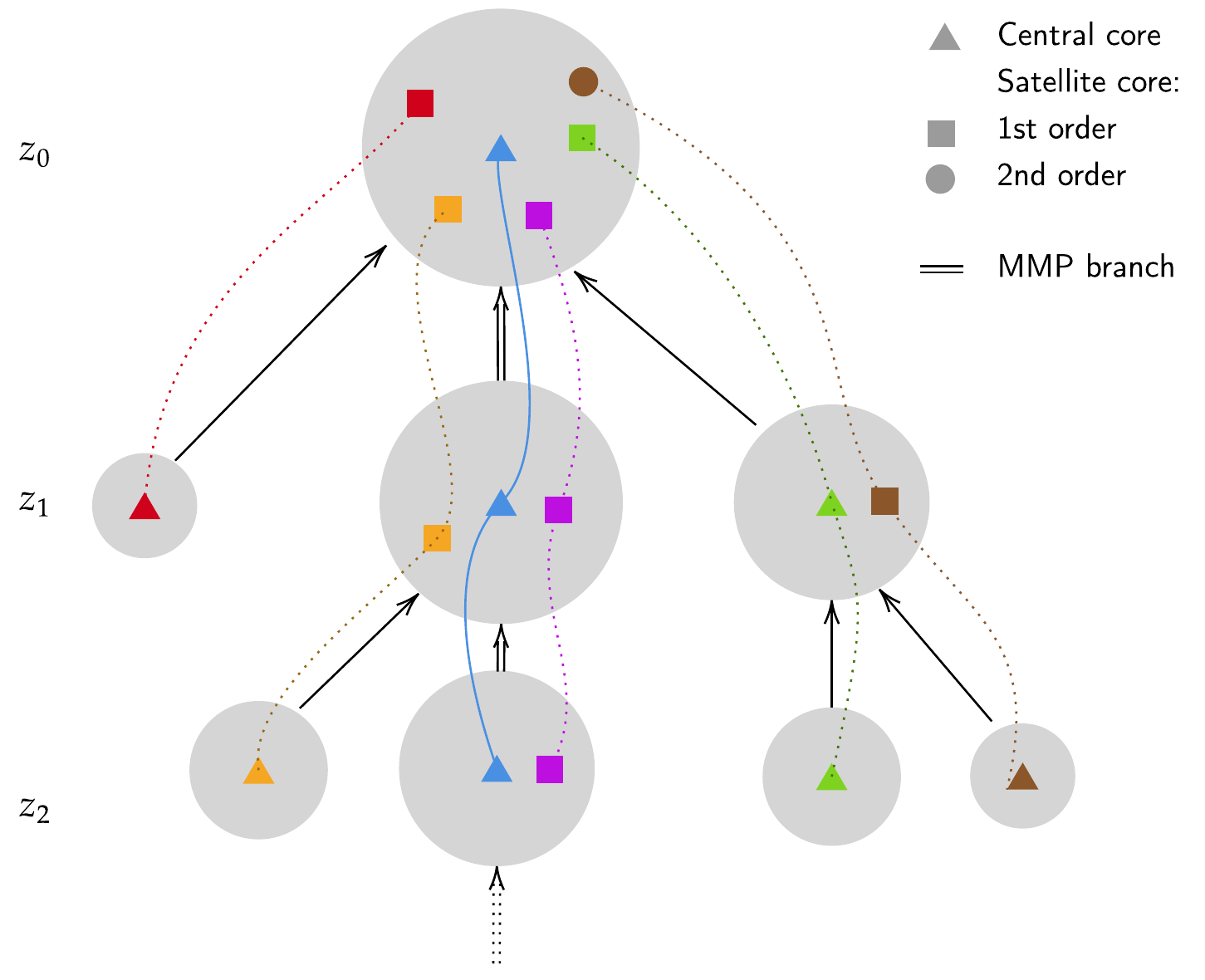}}
\caption{\label{fig:cores_mergertree} Example halo merger tree showing the types of cores associated with halos (gray disks). Central cores (triangles) are identified when an FOF halo is found above a preset core host halo mass threshold. As halos merge, the central core of the most massive progenitor (MMP) is connected to the measured central of the descendant. Cores of less massive progenitors are designated as satellites. The `order' of a satellite traces the number of previous mergers away from a MMP; cores that merged directly into a MMP halo are labeled first-order satellites (squares), cores inherited from one generation above an MMP merger are second-order satellites (circles), etc. This ordering of cores is analogous to substructure hierarchies described in the literature (e.g., \citealt{giocoli10, jvdb14, JvdB2016a}).} 
\end{figure}

There are many complications in merger tree construction, with a number of papers devoted to the subject (e.g., \citealt{fakhouri08}, \citealt{behroozi13}, \citealt{sublink} and \citealt{han18}). 
Finite mass resolution, for example, can cause stochastic threshold effects of the smallest halos, which are repeatedly found, lost, and re-found at masses near the halo minimum mass threshold. We have mitigated this complication by setting our halo finding thresholds to be below our minimum host halo masses in our trees, effectively pruning out halos of masses below our tracking resolution. The smallest resolved halos in our merger trees are recorded by the core host halo mass in Table~\ref{tab:sims}. 

Merger tree construction is further complicated by halo `splitting' events, where nearby (or flyby) halos are temporarily considered as one object (over-linked by an FOF finder, for example) erroneously indicating a merger event, and are later discovered to be different separate (split) objects. To avoid splitting effects, we have found that the construction of merger trees is best carried out in post-processing, starting from the final time step of interest (in our case at $z=0$) and walking backwards in time. Thus, we are ensured that every merger tree will be rooted by one final descendant, and we employ an artificial halo `fragmentation' procedure to account for halo splitting events. Briefly, when we detect a halo progenitor with multiple descendants (indicating the halo split), we fragment the progenitor into separate halos defined by the overlapping particles of each descendant. We then assign to each fragment a portion of the original progenitor mass based proportionally on the descendant mass, thereby conserving total mass (see ~\citealt{rangel17} for details). The resulting merger trees are devoid of splitting events, and consistently track the formation history of each halo.  

In summary, we efficiently record the merger history of all halos to form a connectivity tree catalog that consistently tracks all mass that eventually ends up in the final halos of interest at $z=0$. For each object within the tree, we further store all of the relevant halo catalog properties (mass, velocity, shape, etc.), with adjusted values used for fragmented halos when needed. 

\subsection{Core Catalogs}
\label{subsec:corecat}

By combining the information from core tracking, halo properties, and merger trees, one can construct core catalogs that contain the evolution history of each core from its birth to the final timestep. Core catalogs are analogous to subhalo merger trees and can track the substructure evolution of halos. A detailed description of core catalog construction is given in ~\cite{korytov}. Figure~\ref{fig:cores_mergertree} shows an example of a halo merger tree that also includes core information. Each halo in the merger tree contains one central core, as well as satellite cores that are inherited by merged progenitor halos. 

Core catalogs are constructed by traversing merger trees forward in time. For every halo in the tree, we extract the central particles to determine its central core. When halos merge, the central of the most massive progenitor (MMP) is connected to the measured central of the descendant halo. The other less massive progenitors contribute their own centrals to the descendant, which in turn are labeled as satellite cores. The satellites will trace the dynamical substructure evolution of these progenitor halos.  Moreover, if any progenitor contains satellites from previous mergers, these objects are also carried into the descendant halo, maintaining the hierarchical information of previous structure formation. 

Analogous to the subhalo hierarchies typical of those described in the literature, e.g., \cite{giocoli10, jvdb14, JvdB2016a}, one can assign a satellite `order' based on the generation a central was merged and converted to a satellite. Centrals that merged directly into an MMP are considered {\it first-order} satellites. Centrals merging into halos containing a first-order central are labeled {\it second-order}, and so on, as illustrated in Figure~\ref{fig:cores_mergertree}. 

As all core particles are stored in the simulation output data since birth, we can accurately update satellite trajectories within halos by extracting the updated core particle positions and velocities at each snapshot.  Moreover, every satellite core carries the infall mass and redshift of its host halo when it merged. These quantities will inform the mass-loss model for the cores that we discuss in Section~\ref{sec:massmodeling}. 

\begin{table*}[t] 
\begin{center}
\caption{Host halo mass bins for model fitting (HM)}
\begin{tabular}{c|c|c|c|c}
$\log \left[ M/ \left(h^{{-1}}\mathrm{M_\odot} \right) \right]$ & Host halo count & Fitting range in $\log(m)$      & $(\mathcal{A},\zeta)_{\min\left(\langle\chi_{\mathrm{dof}}^2\rangle\right)}$ & $\min\left(\langle\chi_{\mathrm{dof}}^2\rangle\right)$ \\\hline\hline
{[}12.0, 12.5{]}                                                & 39 505 (33 020)           & {(}$\log(100m_{\mathrm{p,SV}})$, 11.9{]} & (0.9, 0.001)                                         & 2.70                          \\
{[}13.0, 13.5{]}                                                & 4738 (2368)           & {(}$\log(100m_{\mathrm{p,SV}})$, 12.5{]} & (1.5, 0.175)                                         & 3.41                          \\
{[}14.0, 14.5{]}                                                & 348 (21)             & {(}$\log(100m_{\mathrm{p,SV}})$, 13.0{]} & (1.1, 0.07)                                        & 1.77
\end{tabular}\label{tab:fittingbins}
\end{center}
\begin{tablenotes}
    \item 
    \small Mass bins listed in this table are used to optimize the mass-loss modeling parameters. 
    Main values of the second column correspond to $z=0$, and the values in parentheses correspond to $z=1$. $m_{\mathrm{p,SV}}$ is the particle mass of the SV simulation. Results are shown in Figures~\ref{fig:paramexp} and \ref{fig:1sigma}. The three bins are also
    used for the verification test in Figure~\ref{fig:resolution_test}.
\end{tablenotes}
\end{table*}

\section{Mass Modeling}
\label{sec:massmodeling}
Developing and verifying a viable mass-loss model for core-based substructure is a necessary step in the creation of a usable proxy for subhalo merger trees. As described in Section~\ref{subsec:corecat}, satellite cores are substructure tracers of merged halos. We aim to model the mass-loss of a satellite object during its post-merger evolution. In this section, we describe our mass-loss model implementation, compare our results to subhalo finder measurements to determine appropriate fitting parameters, and report on resolution tests. 

The primary metric we use to assess the accuracy of the mass-loss modeling approach is the subhalo mass function (SHMF), given by $\mathrm{d}n/\mathrm{d} \log(m/M) (z)$, which is a measure of the number of subhalos (of mass $m$) within a host halo (of FOF mass $M$\footnote[2]{We use the FOF mass for binning, not the fragment mass (defined in Section~\ref{sec:mergertrees}) to be consistent with the subhalo finder measurements that are recorded using FOF mass.}) at some redshift $z$. Unless otherwise specified, the SHMF bins are normalized by the total number of hosts, as well as the bin width $d\log(m/M)$. The reported error bars are determined by the statistical error of each bin. We compare the SHMF of our mass-evolved satellite cores with the corresponding mass function generated from our subhalo finding implementation described in Section~\ref{sec:SHF}. 

\subsection{Subhalo Mass Modeling}
\label{sec:SHMM}

The approach we choose to follow was introduced by \cite{vdB2005} and uses a power-law to represent the average subhalo mass-loss rate. Their model was motivated by the study of idealized isolated circular subhalo orbits. The mass-loss rates of subhalos in N-body simulations have been measured by, e.g., \cite{giocoli08}, who found good agreement with the power-law model proposed by \cite{vdB2005}. Additional studies have extended the model \citep{JvdB2016a,vdB16,hiroshima18}. We begin by summarizing the mass-loss equations, in a manner consistent with the notation conventions of \cite{JvdB2016a}.

\begin{figure*}[t]
\centerline{\includegraphics[width=2.42in]{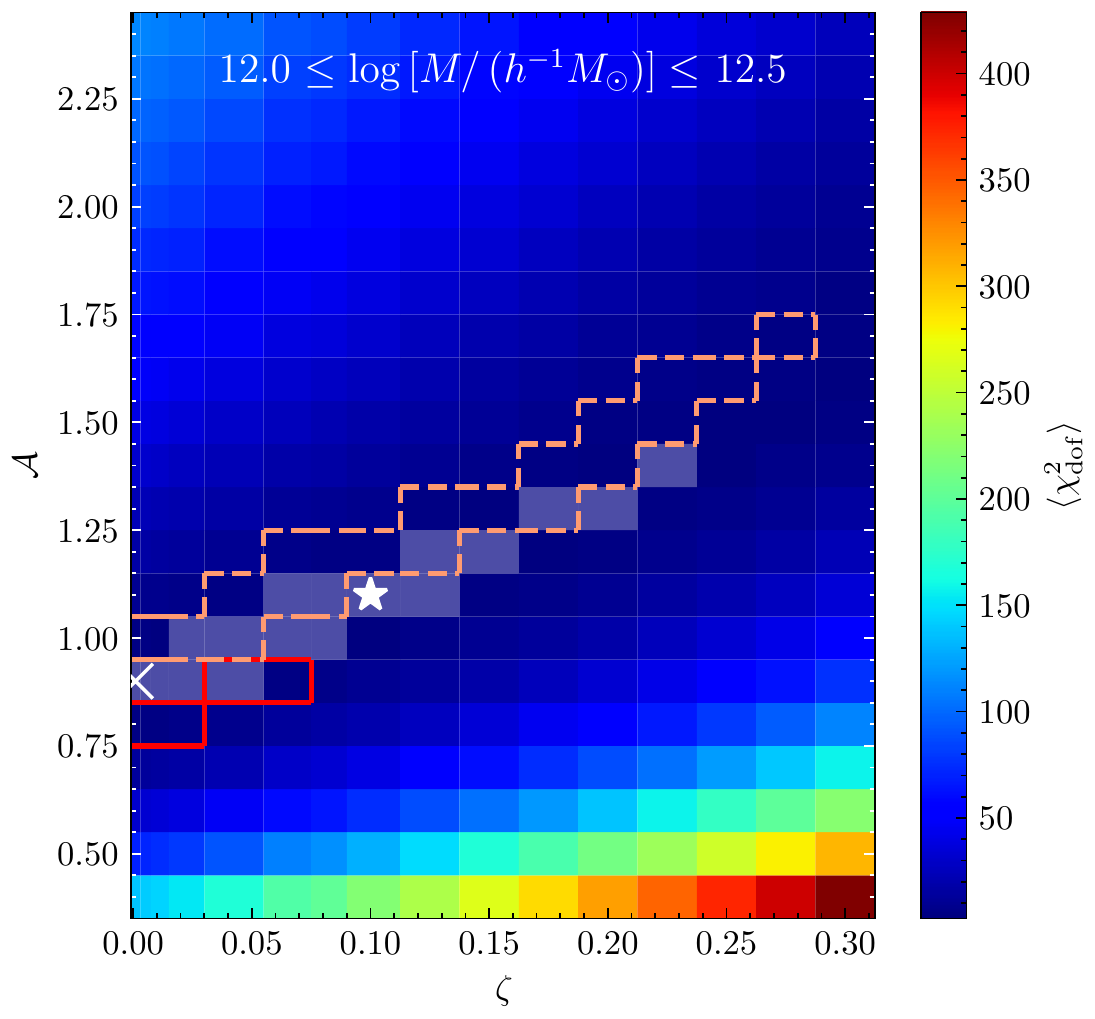}
\includegraphics[width=2.42in]{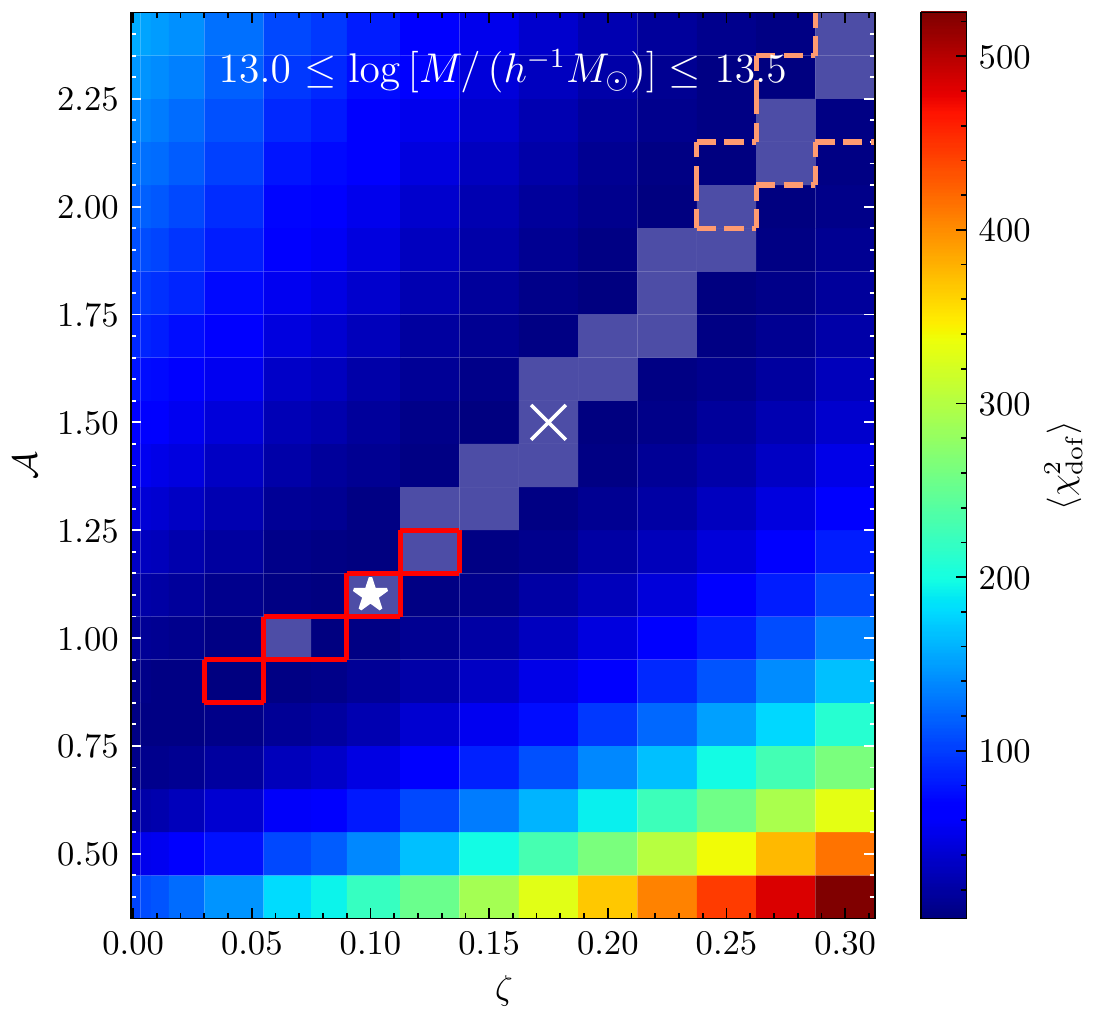}
\includegraphics[width=2.42in]{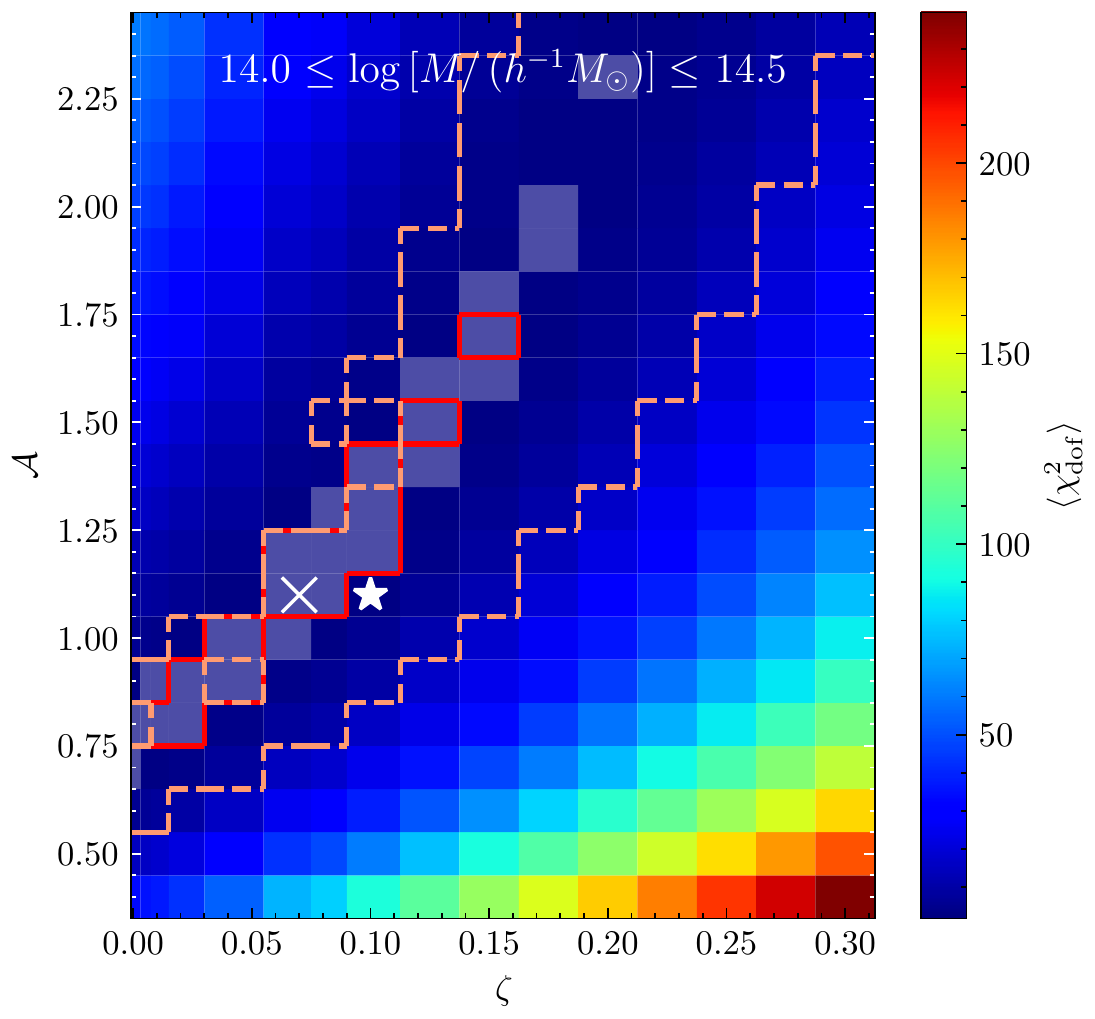}}
\caption{\label{fig:paramexp} Measurements of $\langle\chi_{\mathrm{dof}}^2\rangle$, i.e., the mean of $\chi_{\mathrm{dof}}^2$ over $z=0$ and $z=1$; all results are for the HM simulation. We limit our fitting to cores and subhalos with $m>100m_{\mathrm{p,SV}}$ (where $m_{\mathrm{p,SV}}$ is the particle mass of the SV simulation), and show the ($\mathcal{A}, \zeta$) free parameter exploration for three different host halo mass bins. For each host mass bin, the region in parameter space where $\Delta\langle\chi_{\mathrm{dof}}^2\rangle\le1$ is shaded in white, and the best-fit parameters are marked with a cross. We also outline the region where $\Delta\chi_{\mathrm{dof}}^2(z=0) \le 1$ (red solid line), and the region where $\Delta\chi_{\mathrm{dof}}^2(z=1) \le 1$ (pink dashed line). The mass regions over which we calculate  $\langle\chi_{\mathrm{dof}}^2\rangle$ for the three host mass bins and the resulting best-fit parameters are recorded in Table \ref{tab:fittingbins}. The star indicates our fiducial model parameters ($\mathcal{A}, \zeta=1.1,0.1$) that we use in the analyses that succeed Section \ref{sec:parameterexploration}.
} 
\end{figure*}

The average subhalo mass-loss rate over all orbital configurations is assumed to be given by
\begin{equation} \label{eq:vdB_masslossrate}
    \dot{m}=-\mathcal{A}\frac{m}{\tau_{\mathrm{dyn}}}\left(\frac{m}{M}\right)^\zeta,
\end{equation}
where $M=M(z)$ and $m=m(z)$ are the parent halo mass and subhalo mass at redshift $z$ respectively, and $\mathcal{A}$ and $\zeta$ are free parameters. The dynamical time of the halo is
\begin{equation}
    \tau_{\mathrm{dyn}}(z) = 1.628h^{-1}\mathrm{Gyr}
    \left[\frac{\Delta_{\mathrm{vir}}(z)}{\Delta_{\mathrm{vir}}(0)}\right]^{-1/2}\left[\frac{H(z)}{H_0}\right]^{-1},
\end{equation}
where $\Delta_{vir}(z)$ is the virial overdensity. Following \cite{vdB2005} and \cite{JvdB2016a}, we use the fitting function of \cite{bn1998} for the virial overdensity: 
$\Delta_{\mathrm{vir}}(z) = 18 \pi^2 +82 x - 39 x^2,$ where
$x(z) = \Omega_m(z) - 1$ and $\Omega_R=0$.

If the host halo mass is constant ($\dot{M}=0$),
\begin{equation} \label{eq:vdB_shmassevolution}
    m(t+\Delta t)=
    \begin{cases}
        m(t) \exp(-\Delta t/\tau) &\zeta=0, \\
        m(t)\left[1+\zeta \left(\frac{m}{M}\right)^\zeta \frac{\Delta t}{\tau}\right]^{-1/\zeta} &\text{otherwise}.
    \end{cases}
\end{equation}
The timescale, $\tau(z)$, is the characteristic timescale of subhalo mass loss and is defined as $\tau(z)\equiv\tau_{\mathrm{dyn}}/\mathcal{A}$. The mass-loss model was originally developed to process extended Press-Schechter (EPS) merger trees (see \citealt{Zentner07} for a review of EPS), where the two free parameters were selected to best fit the average SHMF of two $\Lambda$CDM simulations for a host halo mass of $M=10^{15}h^{-1}$M$_\odot$. The decay model was found to be in reasonable agreement with the simulation measurements, and has been studied and refined in a number of follow-up papers (e.g., \citealt{JvdB2016a}). We have found that the simple two parameter model is well suited for our goals and provide some of the implementation details in the following.

For our core mass-loss model, we implement Equation~\ref{eq:vdB_shmassevolution} using core catalogs in place of EPS merger trees. The satellites in the core catalog carry the host halo merger tree information required for the mass-loss model implementation. For each satellite core, we store the simulation timestep immediately preceding infall (merging) of the former host halo, labeled $t_\mathrm{infall}$, in addition to the halo properties at $t_\mathrm{infall}$ (including mass) which we call the satellite's infall properties. We then initialize the model based on the infall mass, and evolve the core mass at each subsequent snapshot using Equation~\ref{eq:vdB_shmassevolution} ($\tau$ is computed at the redshift corresponding to $t+\Delta t$). Note that we use ``core mass'' to refer to the modeled mass of the substructure object associated with the core's position.

The parent mass $M(z)$ used to calculate the mass evolution of a core in Equation \ref{eq:vdB_shmassevolution} is chosen to be the mass of its immediate parent halo (potentially traced by another core with its own evolved mass). 
Explicitly, for cores of order $>1$ (as defined in Section \ref{subsec:corecat}), $M(z)$ is the evolved mass of the parent halo into which the core first fell. For cores with order $=1$, the parent halo mass is simply the host halo mass measured at the given time step; note, we specifically use the fragment host halo mass defined in Section~\ref{sec:mergertrees}, to be consistent with our merger tree construction.
This mass integration approach is similar to the procedure of \cite{JvdB2016a} for evolving the mass of subhalos that can have order greater than one.

A brief remark on initialization for the mass model: for objects that are identified at snapshot $t_{\mathrm{infall}}$ and have merged into a host halo by the next snapshot, labeled $t_{\mathrm{satellite}}$, the merging event could have occurred at any time between $t_{\mathrm{infall}}$ and $t_{\mathrm{satellite}}$. We begin the mass evolution at the average of the two snapshots $(t_{\mathrm{infall}}+t_{\mathrm{satellite}})/2$. Thus, the evolved mass of the core at $t_{\mathrm{satellite}}$ is given by Equation \ref{eq:vdB_shmassevolution}, with $\Delta t=(t_{\mathrm{satellite}}-t_{{\mathrm{infall}}})/2$, $m$ given by the the infall (fragment) mass, and $M$ given by the host halo mass at $t_{\mathrm{satellite}}$. The mass evolution for subsequent snapshots is calculated as described above.

Next, we discuss our approach to find the appropriate fitting parameters for the mass-loss model ($\mathcal{A}$ and $\zeta$) and evaluate the validity of the overall SMACC approach.

\subsection{Parameter Exploration}\label{sec:parameterexploration}

\begin{figure*}[t]
\hspace{0cm}\centerline{\includegraphics[width=6.50in]{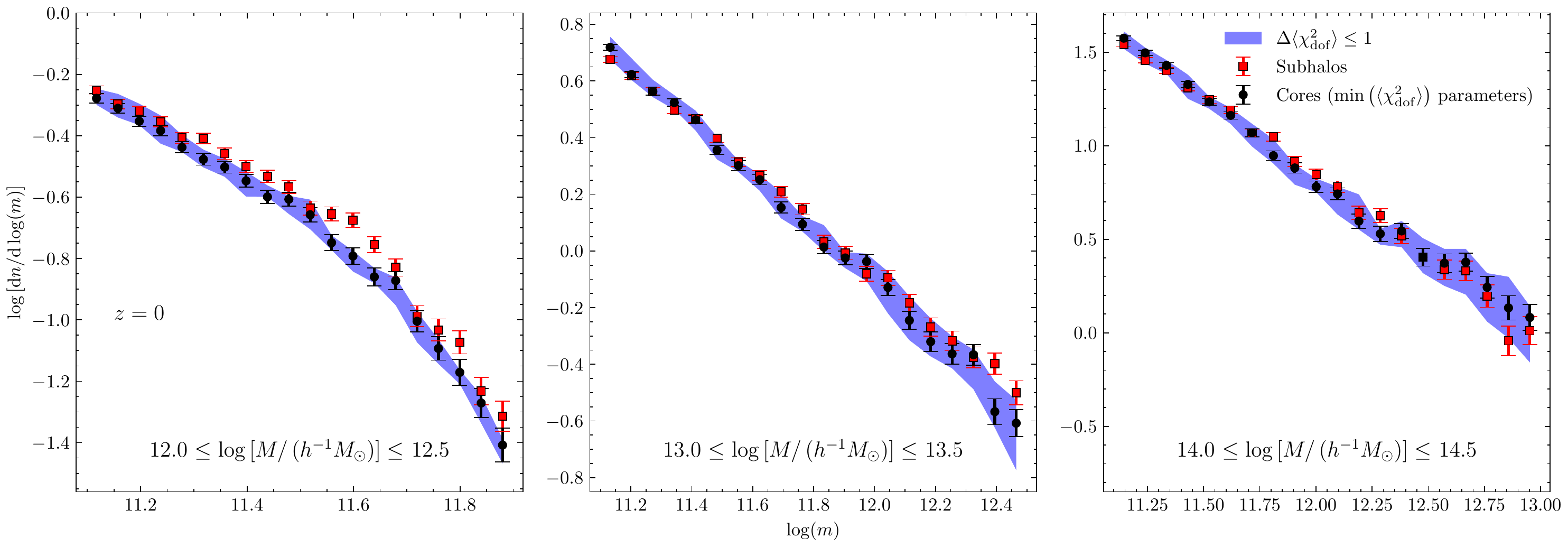}}
\hspace{0cm}\centerline{\includegraphics[width=6.50in]{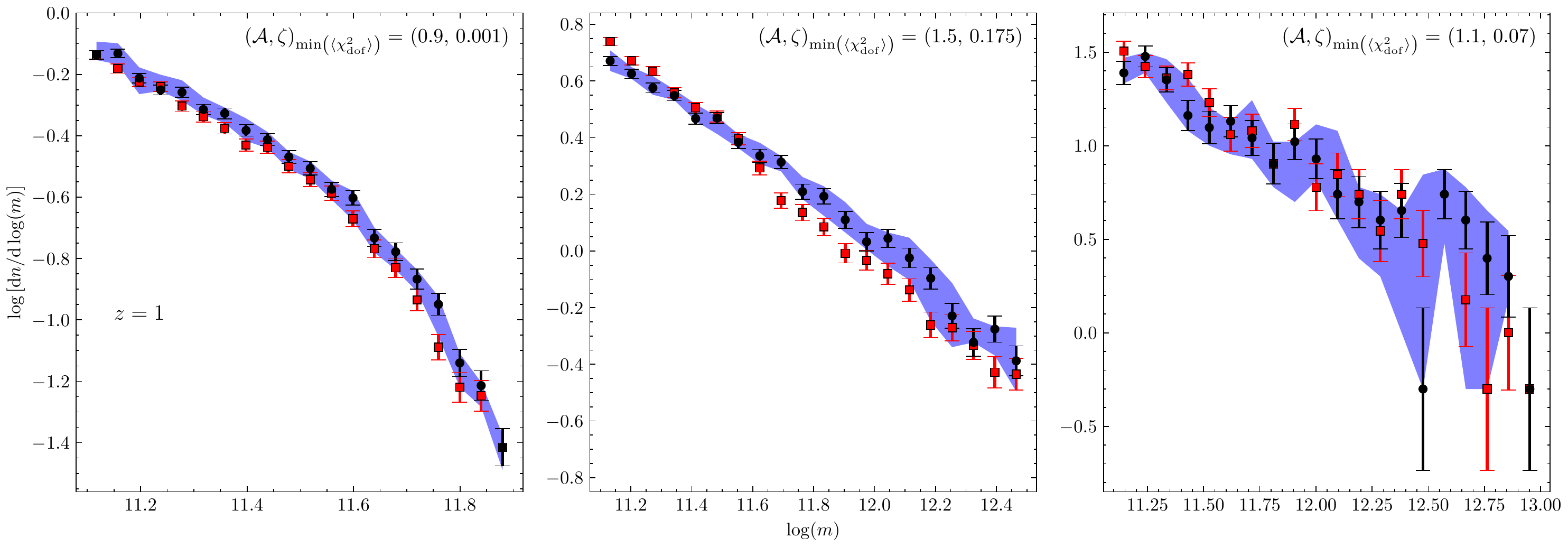}}
\caption{\label{fig:1sigma} Comparison of the subhalo and core mass functions: For three host halo mass bins at $z=0$ (upper panels) and $z=1$ (lower panels) for HM, the shaded region shows the range of core mass functions with the mass model parameters for which $\Delta \langle\chi_{\mathrm{dof}}^2\rangle \le 1$. Also shown is the subhalo mass function, and core mass function using the best-fit (minimum $\langle\chi_{\mathrm{dof}}^2\rangle$) model parameters found for each bin (parameter values are indicated on the lower panels). All mass functions are average mass functions computed in the same mass range as the parameter fitting. See Table~\ref{tab:fittingbins} for the best-fit parameter values, the mass ranges used for fitting, and a listing of the statistics for each bin. The best-fit mass-loss model reproduces the subhalo mass function well; we note that there is a slight redshift dependence (exemplified by the small overall offsets between the model and subhalo measurements in the low mass bin) that is not fully captured by the simplistic model; however, the deviations are small. 
Note that while we show the particular best-fit parameters of each mass bin in this figure, we use the fiducial model parameters ($\mathcal{A}, \zeta=1.1,0.1$) in the analyses that follow Section \ref{sec:parameterexploration}.
}
\end{figure*}

Given the mass-loss model introduced in Section~\ref{sec:SHMM}, we need to determine optimal parameter values of $\mathcal{A}$ and $\zeta$ for Equation~\ref{eq:vdB_shmassevolution}. As the simulations underlying our core catalogs differ from those of~\cite{vdB2005}, we expect a different set of optimal model coefficients. We choose as our fitting criteria the comparison of the predicted SHMF from the core catalogs to that produced by a subhalo finder. We use our HM simulation results at $z=0$ and $z=1$ for this analysis. As shown in Figure \ref{fig:subfinder_comparison} for $z=0$, the HM subhalo finder results agree with the results of SV in the higher subhalo mass overlap region, while the higher mass resolution of HM allows an exploration of lower mass subhalos past the resolution limits of SV.

\begin{figure*}[t]
\hspace{0.cm}\centerline{\includegraphics[width=5.50in]{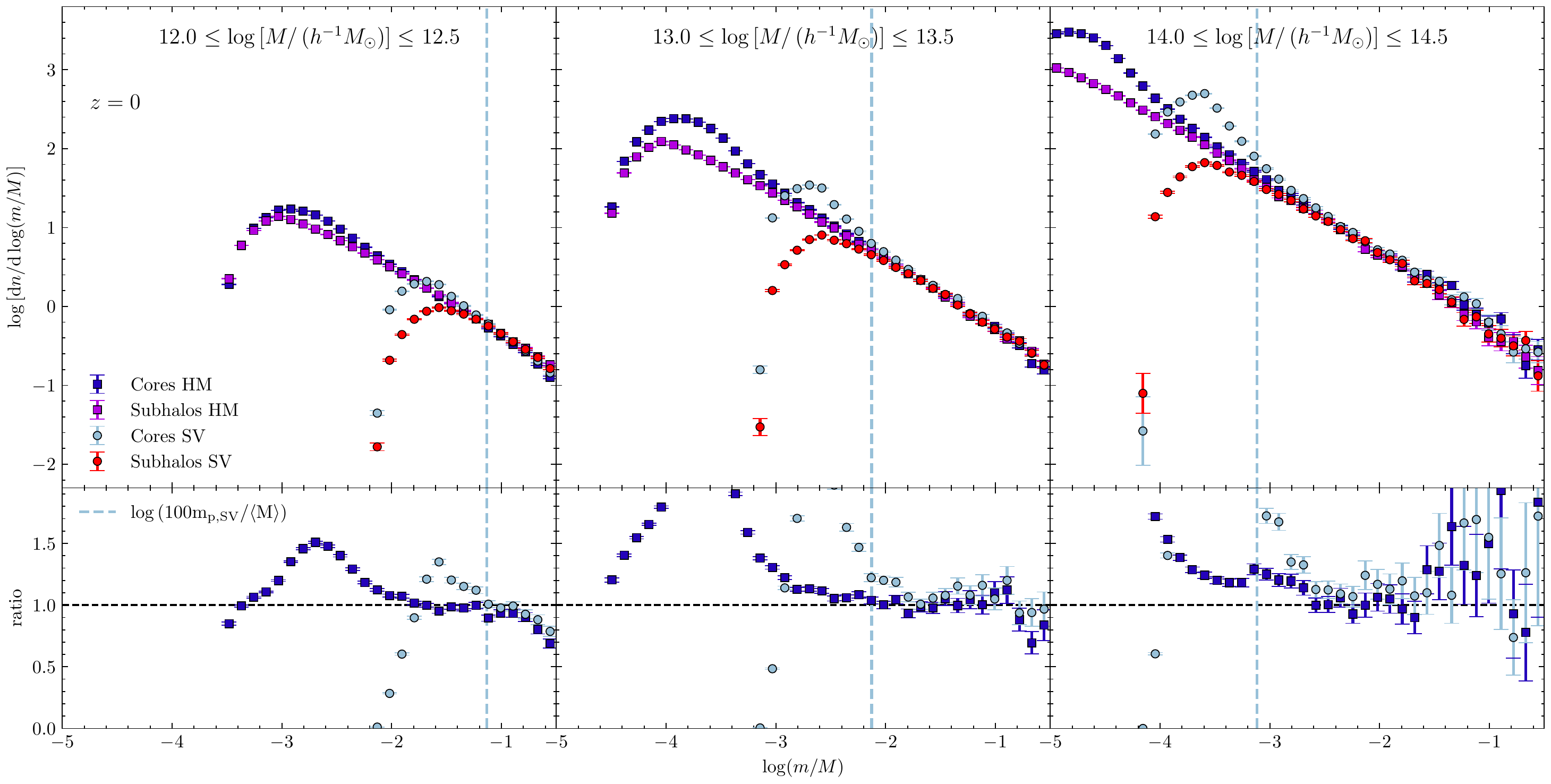}}
\hspace{0.cm}\centerline{\includegraphics[width=5.50in]{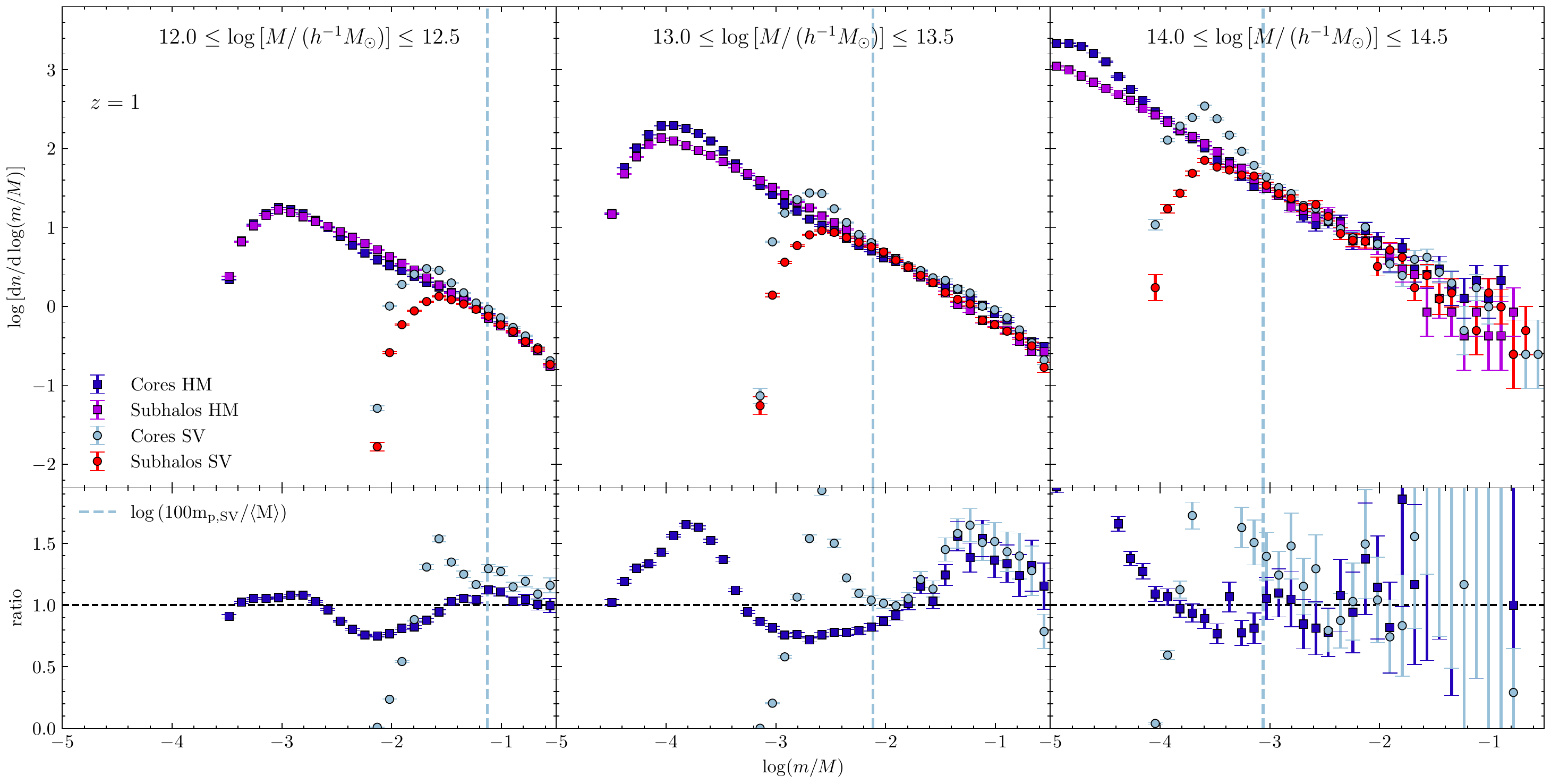}}
\caption{\label{fig:resolution_test} Comparison of the mass model applied to two simulations with different mass resolutions to test resolution effects. Upper panels: The average subhalo and core mass functions at $z=0$ for three host halo mass bins in the HM (squares) and SV (circles) simulations. Subhalos and cores with $m>20$ particles are shown. The HM host bins contain 39505, 4738, and 348 host halos while the SV bins contain 40628, 4797, and 337 halos. The vertical dashed line in each panel approximates the 100 particle mass threshold for SV, above which our model predictions appear to be converged. We additionally measure the ratio of the core mass functions to the HM subhalo mass function, displayed at the bottom of each panel. Lower panels: Same for $z=1$. In this case the HM host bins contain 33020, 2368, and 21 host halos while the SV bins contain 34592, 2418, and 36 halos. }
\end{figure*}

For a given host halo mass bin, we compute the average subhalo mass function $\mathrm{d}n/\mathrm{d} \log(m)$ within $N_{b}=20$ bins of equal width in a specified $\log(m)$ range. Note that we do not normalize $m$ by $M$ for this analysis. We have run our mass model over a grid of the free parameters, spanning
\begin{align*}
    \mathcal{A}=&\{0.4, 0.5, 0.6 , 0.7, 0.8, 0.9, 1.0, 1.1, 1.2, 1.3, 1.4, 1.5, \\
    &1.6, 1.7, 1.8, 1.9, 2.0, 2.1, 2.2, 2.3, 2.4\},\\
    \zeta=&\{0.001, 0.005, 0.01, 0.02, 0.04, 0.07, 0.08, 0.1, 0.125, \\&0.15, 0.175, 0.2, 0.225, 0.25, 0.275, 0.3\}.
\end{align*}
 This allows us to compute the average core evolved mass function $\mathrm{d}n/\mathrm{d} \log(m)$ for the same bins at any point of the grid, $\mathrm{CMF}_i(\mathcal{A}, \zeta)$. We define
 \begin{equation}\label{eq:reducedchisqr}
     \chi_{\mathrm{dof}}^2(\mathcal{A}, \zeta) = \frac{ \chi^2(\mathcal{A}, \zeta)}{N_b-2},
 \end{equation}
 where
 \begin{align}
     \chi^2(\mathcal{A}, \zeta) &= \sum_i\left[\frac{\mathrm{SHMF}_i-\mathrm{CMF}_i(\mathcal{A}, \zeta)}{\sigma_i(\mathcal{A}, \zeta)}\right]^2,\\
     \sigma_i(\mathcal{A}, \zeta) &= \sqrt{\Delta(\mathrm{SHMF}_i)^2+\Delta \left(\mathrm{CMF}_i(\mathcal{A}, \zeta)\right)^2}.
 \end{align}
Here $\mathrm{SHMF}_i$ and $\mathrm{CMF}_i$ are the subhalo mass functions and core-evolved mass functions in bin $i$, respectively 
and
$\Delta(\mathrm{SHMF}_i)$ and $\Delta(\mathrm{CMF}_i)$ are the statistical errors for each bin.
Finally, we compute $\chi_{\mathrm{dof}}^2$ over the grid scan of the parameters at $z=0$ and $z=1$ independently, and measure the mean over the two redshifts,
\begin{equation}
    \langle\chi_{\mathrm{dof}}^2\rangle=\frac{1}{2}\left[ \chi_{\mathrm{dof}}^2(z=0)+\chi_{\mathrm{dof}}^2(z=1) \right].
\end{equation}

The resulting values of $\langle\chi_{\mathrm{dof}}^2\rangle$ are shown in Figure~\ref{fig:paramexp} for three host halo mass bins of the Last Journey-HM simulation. Table~\ref{tab:fittingbins} shows the host halo mass ranges and counts for the three bins, as well as the ranges of $\log(m)$ over which we calculate $\chi_{\mathrm{dof}}^2$ for each redshift. 
Note that we use a mass of 100 particles of the SV simulation ($\log m\sim11.1$) as the fitting range lower bound for all three mass bins; we show in Section \ref{sec:resolutiontests} that this threshold is the approximate resolution limit above which the HM and SV substructure mass functions are mutually converged. The fitting range upper bounds were chosen to be just before the high subhalo masses fall off of the subhalo mass function.

For each host bin, we designate the grid point where $\langle\chi_{\mathrm{dof}}^2\rangle$ is minimized to be the best-fit parameters. This point in parameter space is marked by a cross in Figure \ref{fig:paramexp}. Table \ref{tab:fittingbins} shows the free parameter values and $\langle\chi_{\mathrm{dof}}^2\rangle$ at this point. The light shaded region in each panel of Figure \ref{fig:paramexp} indicates the grid points for which $\Delta\langle\chi_{\mathrm{dof}}^2\rangle\le1$, where 
\begin{equation} \label{eq:Deltachi2}
    \Delta\langle\chi_{\mathrm{dof}}^2\rangle(\mathcal{A}, \zeta) = \left| \langle\chi_{\mathrm{dof}}^2\rangle(\mathcal{A}, \zeta)-\min\left(\langle\chi_{\mathrm{dof}}^2\rangle\right) \right|.
\end{equation}
We also indicate the individual best-fit regions of each redshift in the three panels: the region where $\Delta\chi_{\mathrm{dof}}^2(z=0) \le 1$ is outlined by the red solid line, and the region where $\Delta\chi_{\mathrm{dof}}^2(z=1) \le 1$ is outlined by the pink dashed line. Note $\Delta\chi_{\mathrm{dof}}^2(z)$ for any point in parameter space is defined by Equation~\ref{eq:Deltachi2} if $\chi_{\mathrm{dof}}^2(z)$ is substituted for $\langle\chi_{\mathrm{dof}}^2\rangle$.

In Figure~\ref{fig:1sigma} we show the (shaded) range of the core mass functions at $z=0$ (upper panels) and $z=1$ (lower panels) for model parameters measuring $\Delta\langle\chi_{\mathrm{dof}}^2\rangle\le1$ for the three host mass bins of Table \ref{tab:fittingbins}. We also show the SHMF and best-fit core mass function in the $\log(m)$ fitting range. First, we note that these two functions are in good agreement across all three mass bins shown. Second, the range of allowed models for the two-parameter fit is rather narrow, establishing that the mass-loss formulation provides a well-constrained model for matching cores to subhalos.

By comparing our subhalo and core model results at $z=0$ and $z=1$, we have found that the core mass function best-fit parameters are different for the two redshifts. This indicates a possible redshift dependence in the mass modeling parameters. The redshift deviation is further demonstrated when comparing the mass functions in the lowest mass bin of Figure~\ref{fig:1sigma}; the best-fit core mass function is biased low compared to the subhalo mass function at $z=0$, whereas the opposite is true at $z=1$. This is due to our chi-square optimization over the average of the two redshifts.  Additionally, the varying best-fit parameter values across the host mass bins shown in Table~\ref{tab:fittingbins} imply a host halo mass dependence of the fitting parameters as well. 

For simplicity, we choose $(\mathcal{A},\zeta)=(1.1, 0.1)$ as our fiducial modeling parameters to use across all redshifts and host halo masses for which we compute the mass model. As shown in Figure \ref{fig:paramexp}, this choice of parameters is within the $\Delta\langle\chi_{\mathrm{dof}}^2\rangle\le1$ region for the first two mass bins, and within the $\Delta\chi_{\mathrm{dof}}^2(z=1) \le 1$ region for the high mass bin. Despite the redshift and host halo mass dependence of the model parameters, we show in Figure \ref{fig:resolution_test} that using the fixed parameters, we are able to produce substructure mass functions at $z=0$ and $z=1$ that have good agreement with the subhalo finder results for both HM and SV to our resolution tolerance (see Section \ref{sec:resolutiontests}). These parameters provide a reasonable compromise across the two redshifts and three mass bins examined in this section using HM. (We have verified this result for the SV simulation by carrying out a similar analysis.) The fixed parameters are indicated in Figure~\ref{fig:paramexp} by a star. We will use the fiducial parameters $(\mathcal{A},\zeta)=(1.1, 0.1)$ for all other comparisons.

\subsection{Resolution Tests}
\label{sec:resolutiontests}

Measuring model robustness to significant changes in simulation mass resolutions is important to understand the numerical limitations and tolerances of SMACC.  We expect resolution effects on our core mass model as the particle sets that trace substructure, in addition to the parent halos themselves, are susceptible to
the force and mass resolution limitations of the underlying simulation. Moreover, we expect that our core mass model will have a resolution dependent lower mass limit; above such a threshold, our results would be converged, and represent the minimum mass of substructures we can reliably model. During our measurements, we want to ensure that the resolution mass cuts we employ for the core catalog in the large Last Journey simulation result in a complete set of halo cores equivalent to subhalos measured above the same mass threshold. We remind the reader that a major aim of the Last Journey series is to generate a set of core merger trees that can be used as an alternative to subhalo merger trees as input to  semi-analytic models.

In order to investigate the resolution effects, we use the core catalogs from the Last Journey-SV and HM simulations described in Section~\ref{sec:simulations}, at redshifs $z=0$ and $z=1$. (The mass resolution in the two simulations differs by almost a factor of 30.) We also employ the results from the subhalo finder for this test. Figure~\ref{fig:resolution_test} shows our results for three host halo mass bins (the first three bins listed in Table~\ref{tab:fittingbins}) at both redshifts. For this test, we identified subhalos with over 20 particles and tracked cores at the same resolution. Each of the three panels shows two sets of curves corresponding to the core and subhalo mass functions from both simulations. We found a minimum core mass threshold of $\sim 100$ particles of SV to be sufficient to limit resolution effects of the simulation in our mass model. We mark the mass cut in each panel via the vertical dashed line. 

Collectively, for all host halo mass bins at both redshifts, the results in Figure~\ref{fig:resolution_test} for the high and low resolution simulations demonstrate excellent agreement for the measured core and subhalo mass functions up to the marked mass cut limit. This result is very satisfying as it shows that our results are indeed converged. Moreover, the individual agreement between the subhalo and core mass functions for each simulation demonstrates the accuracy of our optimized mass model. As the parameter tuning discussed in Section \ref{sec:parameterexploration} was carried out on the HM simulation, the matching results of the SV simulation utilizing the same parameters is an indication that our model fitting was robust. Lastly, we note that both simulations exhibit a fall-off in the mass functions after their individual resolution limits are reached. 

As the Last Journey simulation ran with a similar mass resolution to SV, an equivalent threshold of 100 particles correspondingly produces accurate mass evolved substructure measurements. The results of Figure~\ref{fig:massfnLJ} in Section~\ref{sec:results}, demonstrate the quality of measured SHMF in the Last Journey analysis using the reported mass cut. In summary, the resolution tests show the robustness of our approach to mass resolution and allow us to confirm that our chosen mass limit for the final SAM project is valid.

\section{Model Robustness}
\label{sec:validation}

In this section we present results from further investigations of the robustness of the SMACC approach. We compare predictions from SMACC directly with measurements from our subhalo finder, going beyond the SHMF shown in the previous section. We focus on the analysis of the SV simulation since its mass resolution is representative of the main Last Journey simulation. We will discuss effects of the host halo concentration, in addition to results for a one-to-one match-up comparison with subhalos and their associated cores. Lastly, we investigate possible cosmology dependence on our chosen model parameters making use of the AlphaQ simulation. All of these tests are performed on subhalos and cores with evolved masses above a threshold of 100 particles, determined in Section \ref{sec:resolutiontests} to represent sufficiently resolved objects. Our findings provide additional evidence that SMACC enables the use of core catalogs to act as a substitute for subhalo merger trees.

\subsection{Host Halo Concentration Effects}
\label{sec:Concentration}

\begin{figure}[b]
\hspace{0.13cm}\centerline{\includegraphics[width=3.0in]{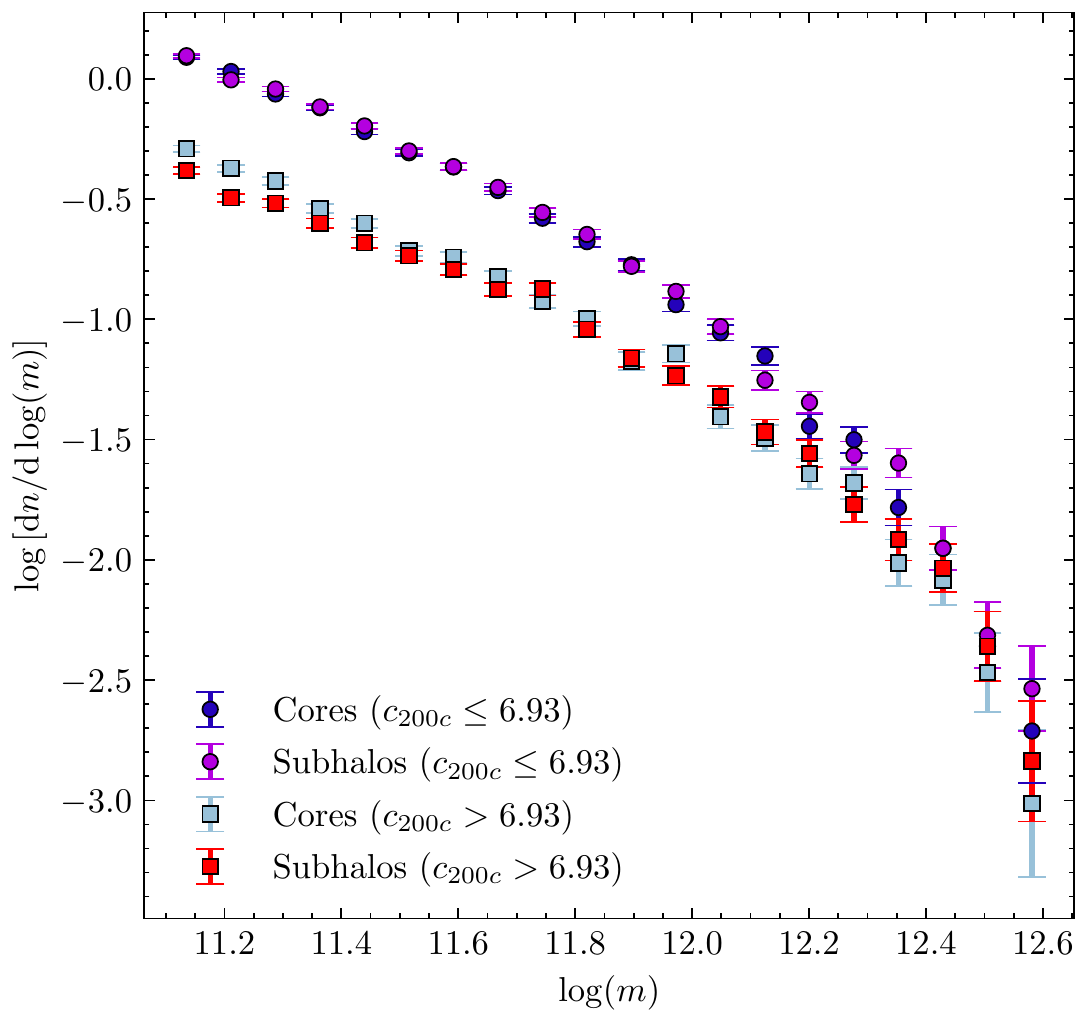}}
\caption{\label{fig:concentration} Average subhalo and core mass functions of the SV simulation at $z=0$ computed for two host halo concentration bins. Note that in this case the $x$-axis is the substructure mass (subhalo and modeled core masses), and that we only consider subhalos and cores which have a mass corresponding to over 100 particles. Host halos in the mass range $12.0 \le \log \left[ M / \left(h^{{-1}}\mathrm{M_\odot} \right) \right] \le 13.0$ are divided into two bins based on the halo concentration $c_{200c}$. The first half $c_{200c}\le\mathrm{Med}(c_{200c})$ and second half $c_{200c}>\mathrm{Med}(c_{200c})$ bins contain 27058 and 27057 halos, respectively, where the host halo median concentration $\mathrm{Med}(c_{200c})=6.93$. Note that our host halos have an FOF mass definition, while their concentrations are derived from the corresponding spherical overdensity halos with $\Delta = 200$ (with respect to the critical density). About 1.5\% of the 54,914 FOF halos within the mass range are removed from our analysis because they are not well-fit by a spherical profile.
} 
\end{figure}

\begin{figure*}[th]
\centerline{{\includegraphics[width=2.42in]{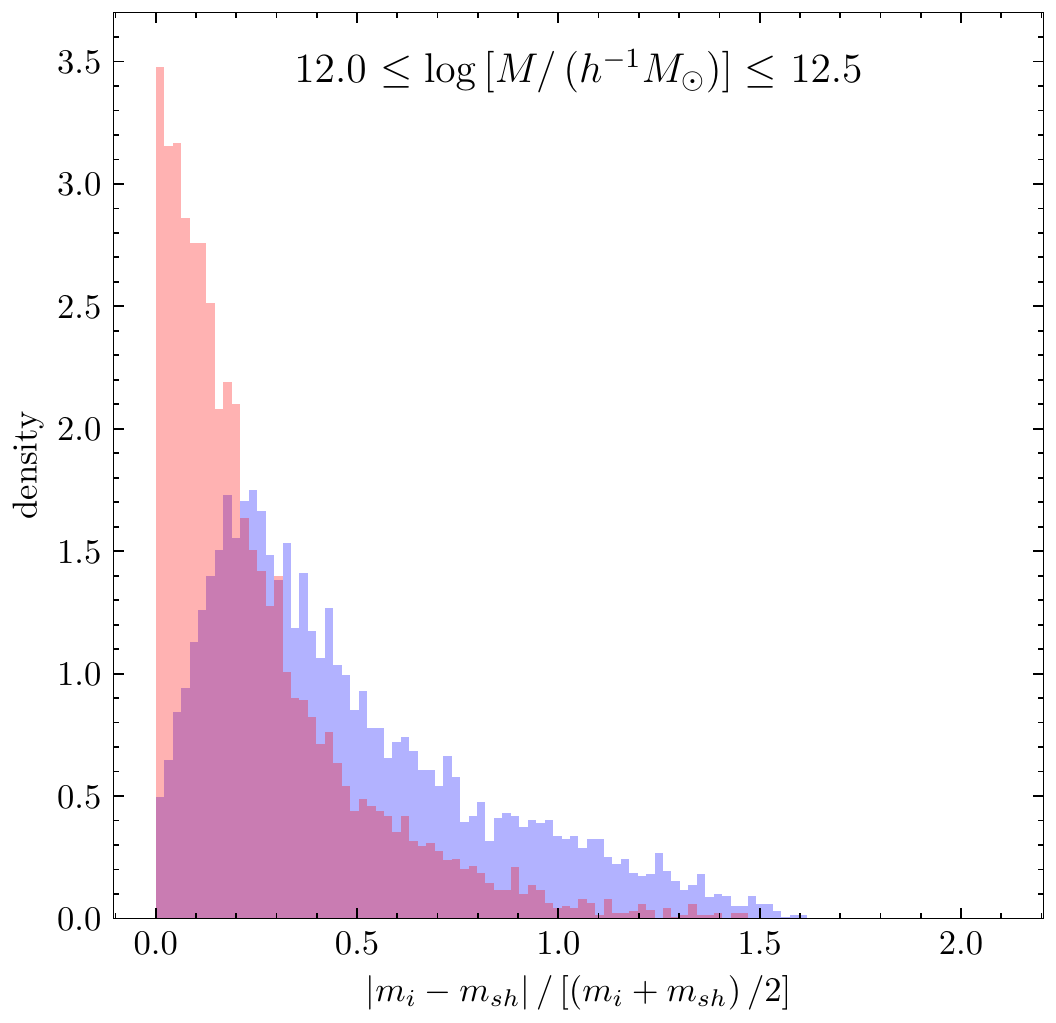}}
{\includegraphics[width=2.42in]{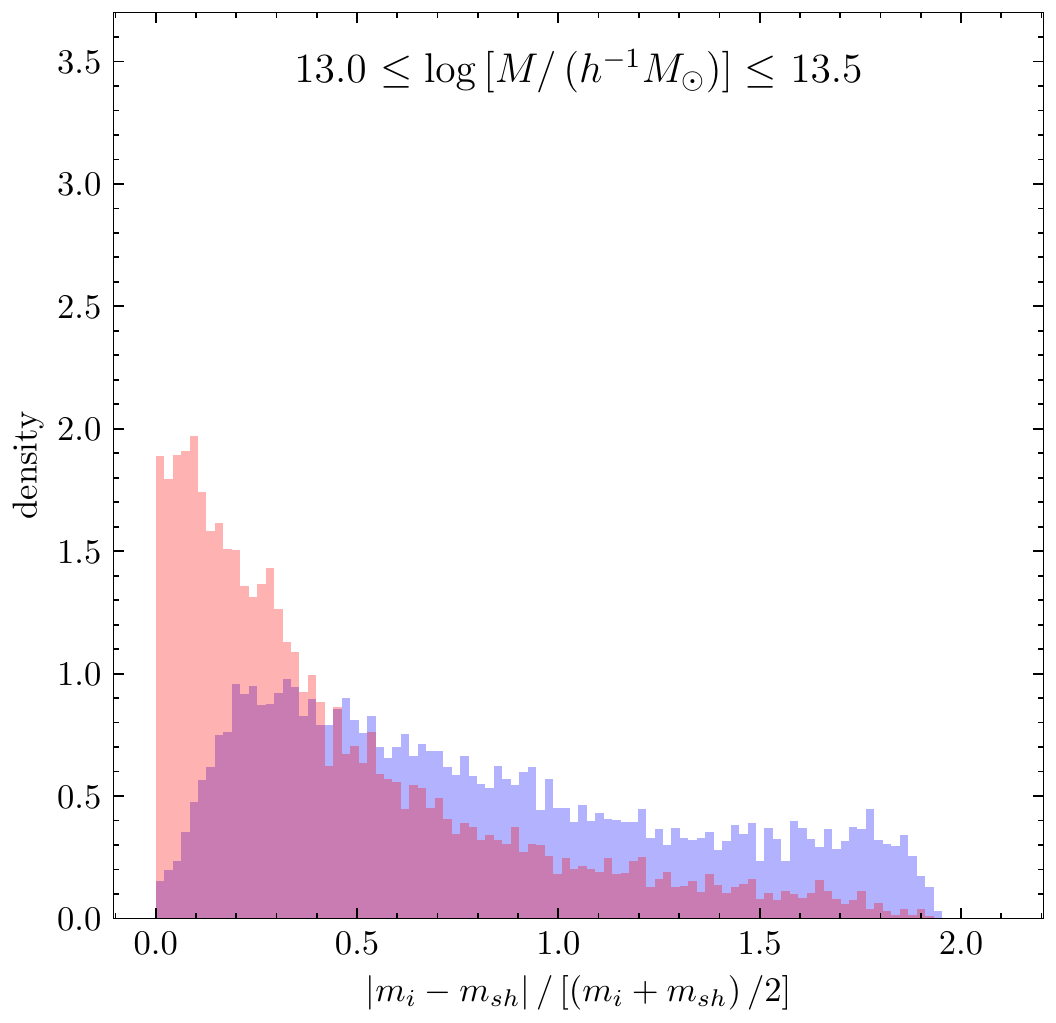}}
{\includegraphics[width=2.42in]{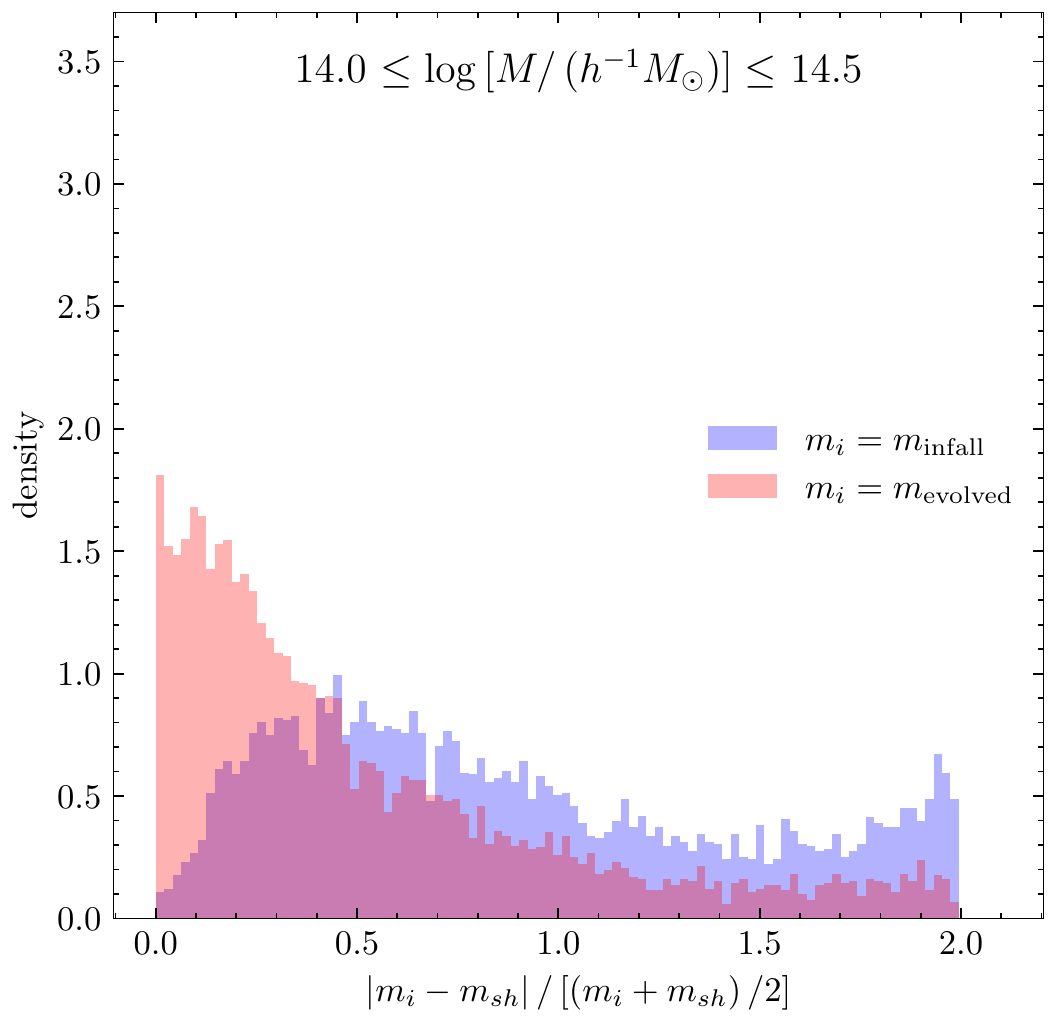}}}
\caption{\label{fig:spatialmatching_reldif} Distribution of the relative difference in mass of spatially matched subhalos and cores in SV. The subset of objects shown are subhalos matched to the most massive core within a search radius of $2r_{sh,vir}$ at $z=0$, where we only consider subhalos and cores with $m>100$ particles for the matching. For the three host mass bins shown, the fractions of matched subhalos (i.e. subhalos that have at least one core above the mass cut within the search radius) are 74\% (left panel), 85\% (middle panel), and 90\% (right panel) of the total subhalos above the mass cut. The relative difference between subhalo mass and core evolved mass (both at $z=0$) is shown in red, where the fiducial model parameters were used to evolve core mass. 
The relative difference between subhalo mass (at $z=0$) and core infall mass (i.e. the mass of halo before merging) is shown in blue.} 
\end{figure*}

\begin{figure*}[ht]
\centerline{{\includegraphics[width=2.42in]{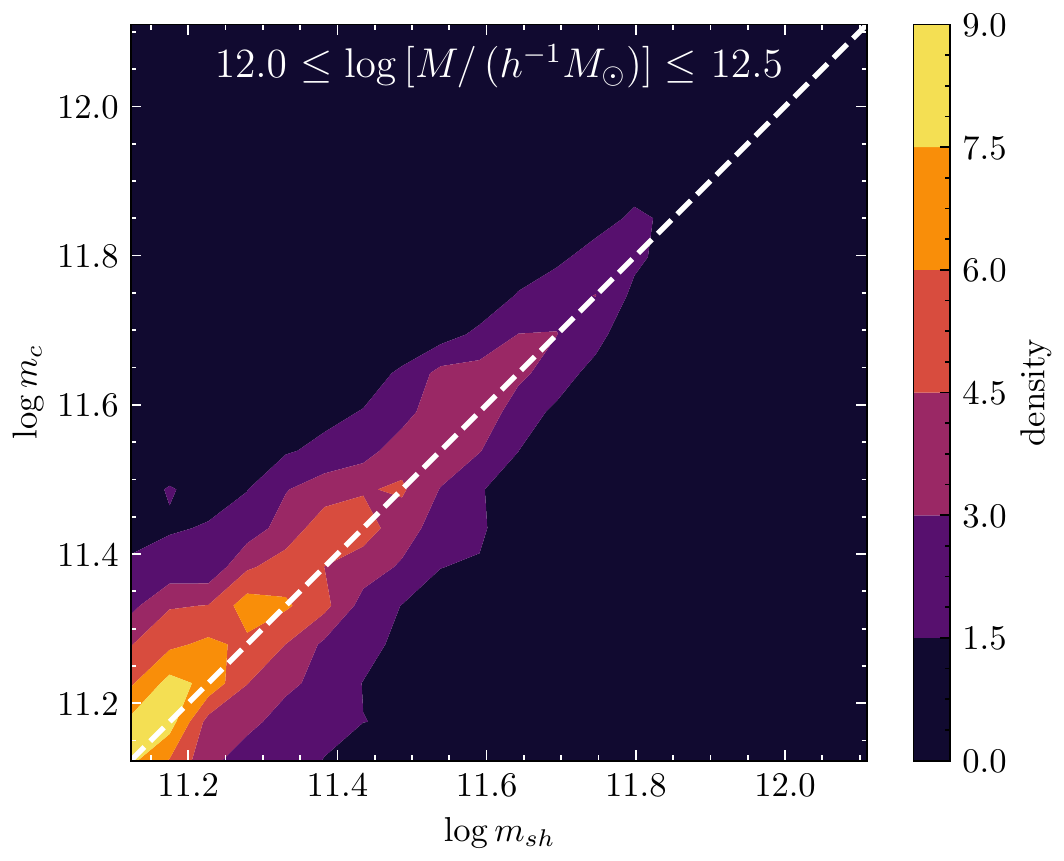}}
{\includegraphics[width=2.42in]{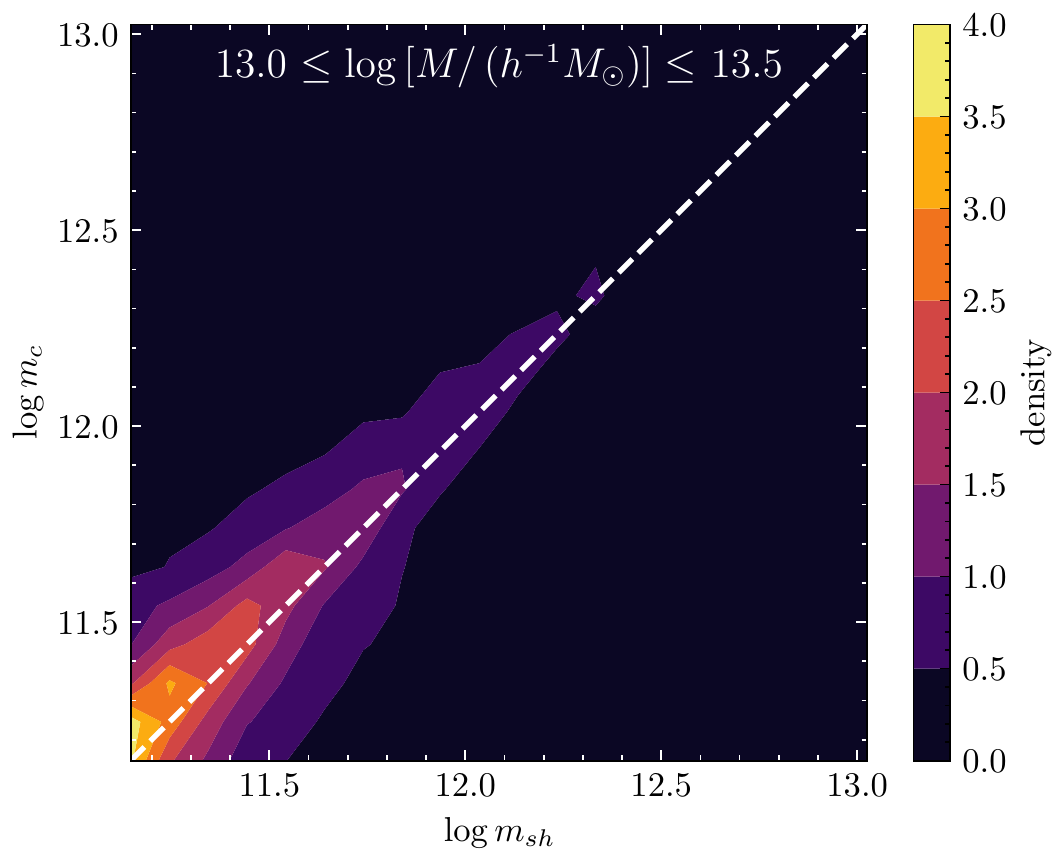}}
{\includegraphics[width=2.42in]{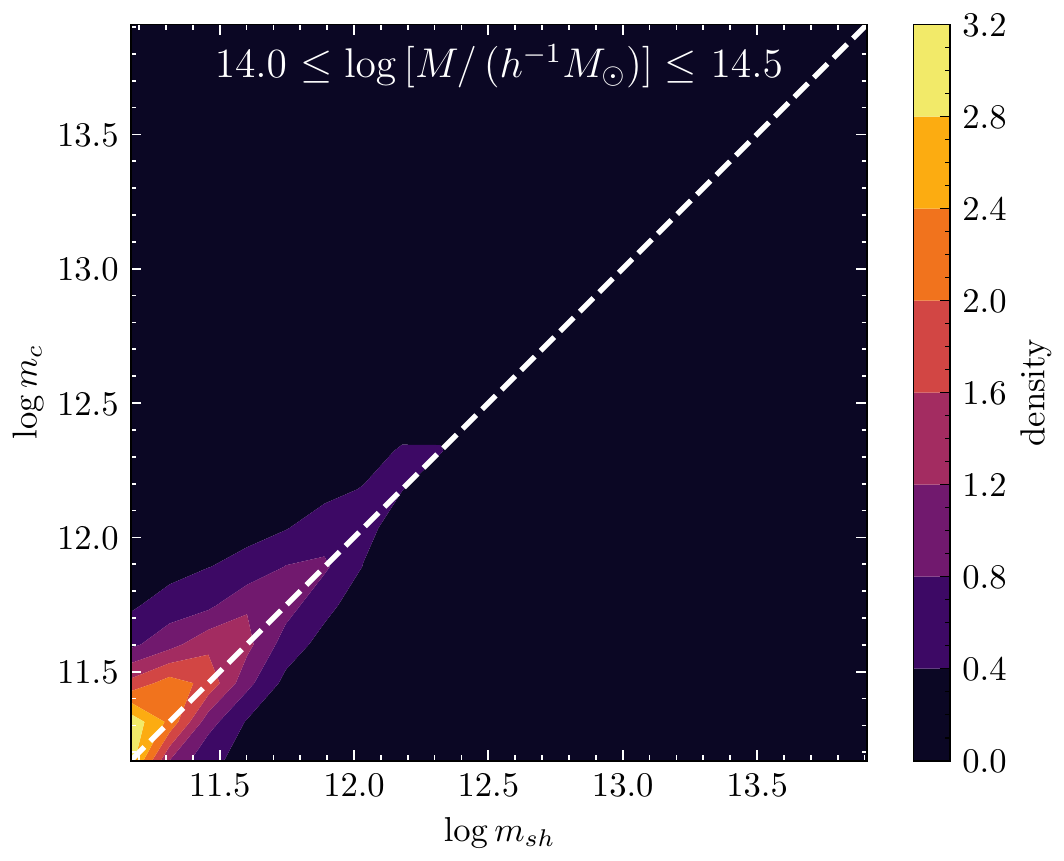}}}
\caption{\label{fig:spatialmatching_contour} Mass comparison contours at $z=0$ of spatially matched subhalos and cores in SV with $m>100$ particles as in Figure~\ref{fig:spatialmatching_reldif}. Subhalo mass $m_{sh}$ is plotted against core evolved mass $m_c$ for three host halo mass bins.
The dashed diagonal lines indicate $m_c=m_{sh}$.}
\end{figure*}

The SMACC approach makes the strong assumption that the core mass $m$ and the ratio with the host halo mass $m/M$ are the only ingredients needed to reliably predict the mass-loss evolution of substructures inside a host halo. However, structure formation is a complex process, and the environment in which a halo formed and grew, in addition to other halo properties such as formation time, surely influence substructure evolution. In order to capture some of this additional information, we employ the host halo concentration as a simple proxy for estimating these effects. The concentration does depend on the environment (see, e.g., \citealt{wechsler02} for a detailed discussion) and therefore will provide a good indication for the robustness of our approach. For example, tidal forces will be stronger in higher concentration halos, so one would expect more subhalo mass loss in this population. The concentration also carries information about the halo's evolution history and local environment. Within a given mass range, halos that have lived a quiet life will have higher concentrations. This implies, once again a lower subhalo mass function, because the subhalo or core mass ``feeding function'' for these halos will likely be smaller.

To investigate the range of possible effects, we partition the halo population by sorting halos within a given mass range into two concentration groups.
We restrict our investigation to a mass bin of 10$^{12}h^{-1}$M$_\odot$ to 10$^{13}h^{-1}$M$_\odot$ for the host halo. While the host halo mass is given by the FOF, $b=0.168$ mass, the concentration is measured using the spherically averaged density profile, and the associated spherical overdensity mass, M$_{200c}$. For each FOF halo we determine the potential minimum and grow a sphere around it, measuring the radial density profile and fitting a Navarro-Frenck-White (NFW) functional form \citep{nfw1,nfw2} to it. We first determine the mass and then fit for the halo concentration. The NFW profile is given by
\begin{equation}
\rho(r)=\frac{\delta\rho_{\rm crit}}{(r/r_s)(1+r/r_s)^2},
\end{equation}
where $\delta$ is a characteristic dimensionless density and $r_s$ is the scale radius. The halo concentration is defined via $c_\Delta=r_\Delta/r_s$, where we take $\Delta$ to be the overdensity with respect to  the  critical  density  of  the  Universe, $\rho_{\rm crit}= 3H^2/8\pi G$. The radius, $r_\Delta$, is reached when the enclosed mass, $M_\Delta$,  equals the volume  of  the  sphere  times  the  density $\Delta\rho_{\rm crit}$. We  measure masses and concentrations for $\Delta= 200$, corresponding in turn to $c_{200c}=R_{200c}/r_s$. A more detailed discussion of our approach can be found in \cite{child} where alternative methods for concentration measurements are discussed as well.

Figure~\ref{fig:concentration} shows the substructure mass functions for host halos in two different concentration ranges of $c_{200c}\le\mathrm{Med}(c_{200c})$ and $c_{200c}>\mathrm{Med}(c_{200c})$, where the median concentration of the host halos in the mass bin is $\mathrm{Med}(c_{200c})=6.93$. As in the previous tests, we show both the subhalo mass function and the core mass function. We consider only subhalos and cores which have a mass over the resolution threshold of 100 particles. First, we note that the mass function for the high concentration halos is increasingly suppressed, compared to that for the low concentration halos, as the substructure mass decreases. This is due to the connection of concentration and halo age, the more concentrated halos are older and therefore substructures have had more time to either merge into the central, high-density region or be disrupted. This finding is in agreement with previous results (see e.g., \citealt{Gao04,Zentner05,wechsler06}). Second, the agreement between the evolved core masses and the subhalo masses in both concentration bins is still excellent. This result confirms that the SMACC approach is robust against different halo formation histories for modeled masses above our resolution threshold.

\subsection{Subhalo-Core Spatial Matching}
\label{sec:CoreSHMatch}
The mass functions used thus far for comparisons between subhalos and cores characterize agreement in overall mass distributions.
We have shown that the core mass-loss methodology, in the sense of averages (e.g., SHMFs), performs well. The final step is to investigate and confirm that the simple description of allowing cores and their history to be associated to a substructure mass is in fact valid at the level of individual objects. This is clearly a necessary requirement for the core approach to be used with detailed semi-analytic models of galaxy formation.

We now investigate how well the mass model works on a case-by-case basis by comparing the mass of an individual subhalo to its corresponding modeled core. For this analysis, we use our SV simulation results at $z=0$ and consider only subhalos and cores which have a resolved mass over 100 particles. We perform a spatial matching procedure by selecting the most massive core within a search radius of $2r_{sh,vir}$ of each subhalo. We define the virial radius of a subhalo, $r_{sh,vir}$, as the radius corresponding to a uniform sphere of mass equal to the subhalo mass, and density equal to $\Delta_{vir}(z)\rho_c(z)$, where $\rho_c(z)$ is the critical density (note that we consider the subhalo mass to be approximately a virial mass for this calculation). The subsets of subhalos which have at least one core (above the mass cut) within the search radius are 74, 85, and 90\% of the total subhalos (above the mass cut) for three host halo mass bins covering a mass range of $[12.0,12.5]$, $[13.0,13.5]$, and $[14.0,14.5]$ in $\log \left[ M / \left(h^{{-1}}\mathrm{M_\odot} \right) \right]$, respectively.

Figure~\ref{fig:spatialmatching_reldif} shows the distribution of the relative difference in mass of the subhalo and matched core for the three host halo mass bins described above. We show results for the infall mass and the evolved core mass (using the $z=0$ subhalo mass for calculating both relative differences), with the expectation that the evolved core mass should show a significantly improved correspondence, i.e., the distribution should shift significantly towards the origin with an increase in the peak height. This is indeed observed and, in particular, for the smaller host halo mass bin, the agreement is rather good. For the larger host halo masses, the matching will be less precise since there are more subhalos per host halo that could provide an alternative (and more suitable) match. Nevertheless, even for the two higher mass bins, the results show that most evolved core masses are close to the corresponding subhalo mass. In Figure~\ref{fig:spatialmatching_contour} the mass comparison for the same data is shown in a different way to study possible biases in the matched masses. Here we show mass comparison contours for the same set of matched subhalos and cores and three mass bins as in Figure~\ref{fig:spatialmatching_reldif}. The contours align very well on the diagonal, indicating good agreement (with some intrinsic scatter) for the matched subhalo-evolved core mass pairs.

\subsection{Cosmology Effects}
\label{sec:Cosmology}

\begin{figure}[t]
\hspace{0.13cm}\centerline{\includegraphics[width=3.0in]{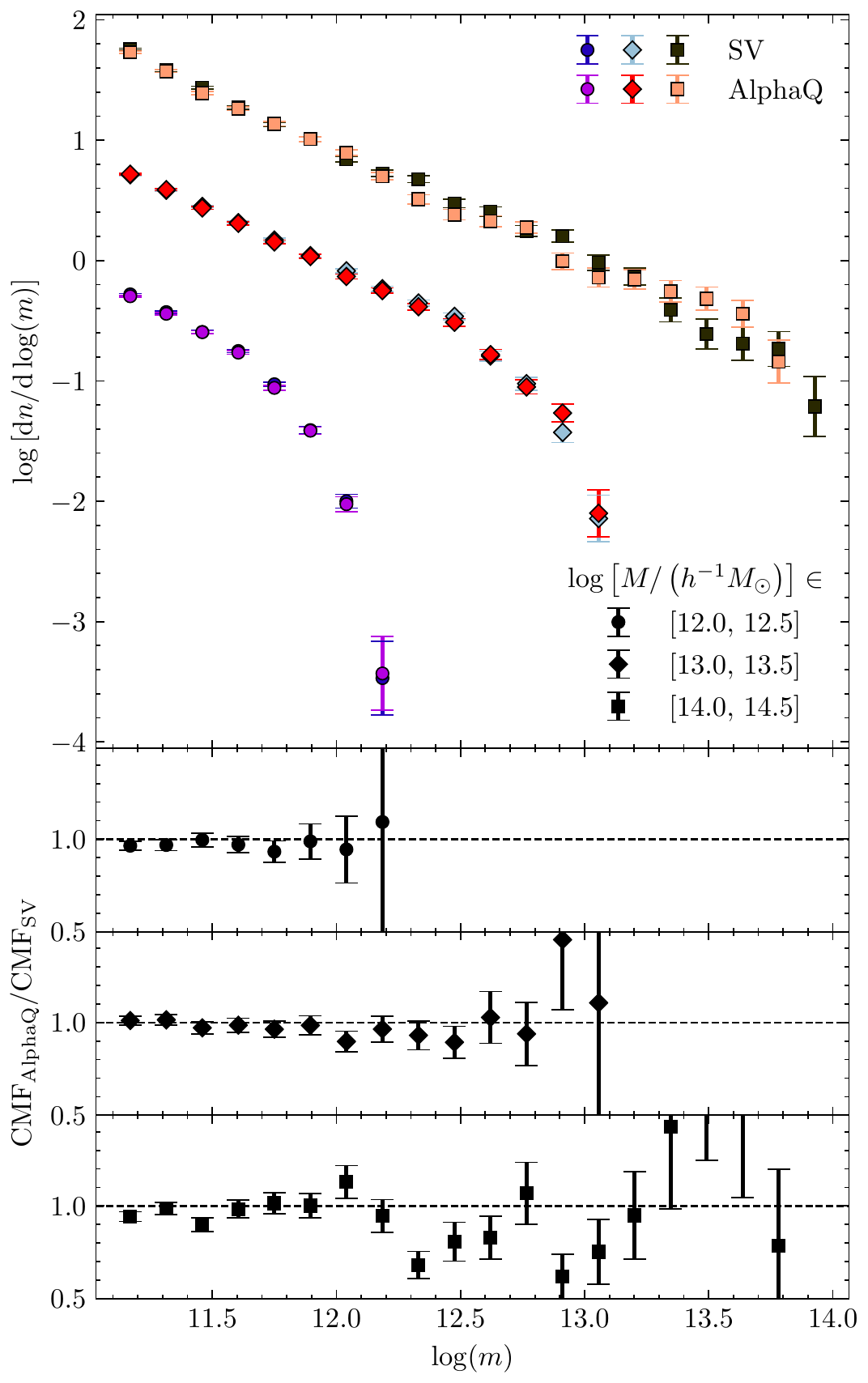}}
\caption{\label{fig:cosmology_test_z0} Comparison of the mass model applied to two simulations with different cosmologies. Upper panels: Average core mass functions at $z=0$ for three host halo mass bins in the SV and AlphaQ simulations. Note that the $x$-axis is the core mass $m$, and that we consider only cores for which $m>100m_{p,\mathrm{SV}}$, where $m_{p,\mathrm{SV}}$ is the particle mass of the SV simulation. The SV host bins contain 40628, 4797, and 337 host halos while the AlphaQ bins contain 37182, 4335, and 286 halos.
The host halo bins cover a mass range of $[12.0,12.5]$ (circles), $[13.0,13.5]$ (diamonds), and $[14.0,14.5]$ (squares) in $\log \left[ M / \left(h^{{-1}}\mathrm{M_\odot} \right) \right]$.
Lower panels: Ratio of the core mass functions shown in the upper panels (AlphaQ to SV for each host halo mass bin).} 
\end{figure}

\begin{figure*}[h]
\centerline{\includegraphics[width=5.4in]{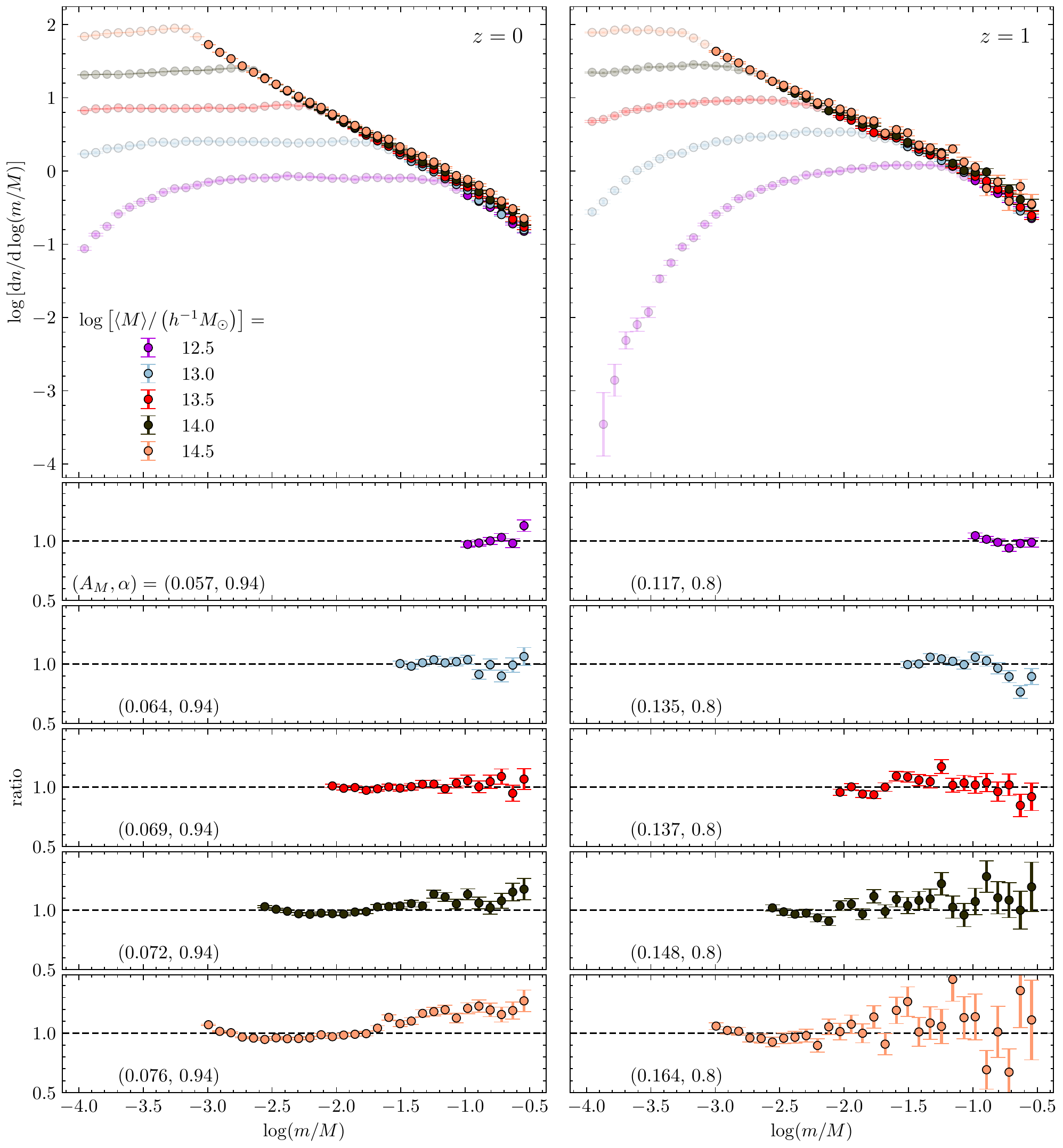}}
\caption{\label{fig:massfnLJ} Results from the main Last Journey simulation for the core mass model.
Upper panels: Average core mass functions at $z=0$ (left) and $z=1$ (right) for five host halo mass bins. (See Table~\ref{tab:hostbinsLJ} for descriptions of the host mass bins.) To illustrate the core mass cut of 100 particles (see Section \ref{sec:resolutiontests}), points for which $\log(m/M)\le\log(100 m_{p,\mathrm{LJ}}/\langle M \rangle)$ have enhanced color transparency, where $m_{p,\mathrm{LJ}}$ is the particle mass of the Last Journey simulation and $\langle M \rangle$ is the average host halo mass of each host halo bin. For each host halo mass bin, we fit Equation~\ref{eq:fittingfn} to the opaque data points of that bin, first fixing $\alpha=(0.94,0.8)$ for $z=(0,1)$ and finding a best fit value for $A_M$.
Lower panels: For each host halo mass bin, the ratio of the opaque section of the core mass function (shown in the upper panel) to the fitting function, for both redshifts. The best fit parameter value for $A_M$ is displayed in each panel. } 
\end{figure*} 

Finally, we investigate the sensitivity of the core mass-modeling approach to a change in the cosmological parameters specifying the simulation. For this test, we employ two simulations -- the first is the Last Journey SV simulation, using the Planck-18 cosmology, while the second, the AlphaQ simulation, is based on a WMAP-7 best-fit model. Both simulations have a volume of ($\sim 250 h^{-1}$Mpc)$^3$ and 1024$^3$ particles. Details of the simulations, including the values for the cosmology parameters, are provided in Section~\ref{sec:simulations}. The main difference between the two cosmologies is reflected in their values for $\Omega_{\rm cdm}$ and $\sigma_8$. Both parameters are larger for the Planck-18 cosmology, leading to enhanced structure formation. In addition, the value for $h$ is smaller in the Planck-18 cosmology compared to WMAP-7. As for our previous tests, we investigate the core mass function for three different host halo mass bins. We apply the same mass model parameters to the two simulations. 

Figure~\ref{fig:cosmology_test_z0} summarizes our findings. We show the core mass functions themselves and the ratio for the two cosmologies in each host halo mass bin. The agreement is very good and shows that the approach is insensitive with regard to cosmology, at least in the family of $\Lambda$CDM models close to the current best-fit cosmology. \cite{vdB16} find similarly that the effect of cosmology on the SHMF is negligible for models close to the current best-fit $\Lambda$CDM model. 

\section{Last Journey Core Mass Function}
\label{sec:results}

Following the description and validation of the SMACC approach above, we now apply the model to the main Last Journey simulation. The simulation evolved 10,752$^3$ particles in a (3400$h^{-1}$Mpc)$^3$ volume, leading to a mass resolution of $m_p\sim 2.721\cdot 10^9 h^{-1}$M$_\odot$. The volume covered is $\sim$2500 times the size of the SV simulation. A detailed description of the Last Journey simulation, including first level analysis results such as matter power spectrum and mass function measurements, is provided in~\cite{LJ1}. 

\begin{table*}[t] 
\begin{center}
\caption{Last Journey host halo mass bins}
\begin{tabular}{c|c|c|c}
$\log \left[ \langle M \rangle / \left(h^{{-1}}\mathrm{M_\odot} \right) \right]$ & $M_{min}$ ($10^{12}h^{{-1}}\mathrm{M_\odot}$)    & $M_{max}$ ($10^{12}h^{{-1}}\mathrm{M_\odot}$)   & Host halo count \\\hline\hline
12.5 (12.5) & 3.163 (3.163)     & 3.163 (3.163)     & 42 640 (32 902)          \\
13.0 (13.0) & 10.000 (10.000)   & 10.005 (10.005)   & 14 303 (8365)           \\
13.5 (13.5) & 31.625 (31.625)   & 31.679 (31.666)   & 10 102 (2929)           \\
14.0 (14.0) & 100.000 (100.000) & 100.674 (100.438) & 10 018 (1093)           \\
14.5 (14.5) & 316.229 (316.245) & 328.172 (331.645) & 10 001 (354)          
\end{tabular}\label{tab:hostbinsLJ}
\end{center}
\begin{tablenotes}
    \item 
    \small Halo masses and counts 
    for the Last Journey simulation. The numbers given in each column correspond to the values for the $z=0$ host halo bins and are followed by numbers in parentheses which  correspond to the values for $z=1$ host halo bins. The bins are used in Fig.~\ref{fig:massfnLJ} for the mass-loss model applied to the large Last Journey simulation. As a consequence of the excellent statistics available for the large Last Journey simulation, we have chosen very narrow mass bins that have a much smaller width than the bins used in the earlier sections for the smaller simulations.
\end{tablenotes}
\end{table*}

The upper panels of Figure~\ref{fig:massfnLJ} show the resulting core mass functions at redshifts $z=0$ (left panel) and $z=1$ (right panel) for five host halo mass bins. As a consequence of the excellent statistics available for the large Last Journey simulation, we have chosen very narrow mass bins that are much smaller in mass width than the mass bins used in the earlier sections for the smaller simulations. Table~\ref{tab:hostbinsLJ} lists the mass bins and the host halo counts for both redshifts. Our chosen parameter values of $(\mathcal{A},\zeta)=(1.1, 0.1)$ were used to evaluate the mass-loss model (see Section \ref{sec:parameterexploration} for a discussion of the optimal parameters). As examined in Section~\ref{sec:resolutiontests}, following the convergence results established with the HM and SV simulations, we impose a mass cut for the core mass for the Last Journey simulation at 100 particles. We consider cores above this threshold to reliably model substructure mass evolution, and they can serve as input for a future SAM project. 
To illustrate this mass cut, points for which $\log(m/M)\le\log(100 m_{p,\mathrm{LJ}}/\langle M \rangle)$ are shown with a higher color transparency ($m_{p,\mathrm{LJ}}$ is the particle mass of the Last Journey simulation; $\langle M \rangle$ is the average host halo mass of each bin). 

For each redshift, to enable comparison of our mass functions with other studies, we fit Equation~\ref{eq:fittingfn} to our core mass function in each host halo mass bin. In \cite{vdB16}, the authors showed that subhalo mass functions determined at $z=0$ using the Bolshoi and MultiDark simulations and the output of two different subhalo finders agree with this fitting form for $\log(m/M)\lesssim-1$\footnote[3]{We note that \cite{vdB16} use the virial mass definition for host halo mass while we use the FOF, $b=0.168$ mass.}. In our results, we consider points for which $\log(m/M)>\log(100 m_{p,\mathrm{LJ}}/\langle M \rangle)$ (i.e., the opaque points shown in Figure \ref{fig:massfnLJ}) when fitting to each mass bin. 

We first fit Equation~\ref{eq:fittingfn} with both $\alpha$ and $A_M$ as free parameters, and found the best-fit slope values at $z=0$ of $\alpha=$ 0.849, 0.953, 0.929, 0.928, and 0.939 for the $\log \left[ \langle M \rangle / \left(h^{{-1}}M_\odot \right) \right]=$ 12.5, 13.0, 13.5, 14.0, and 14.5 bin, respectively. At $z=1$, the best-fit values are 0.883, 0.852, 0.776, 0.770, and 0.797 for the same bins. We empirically selected $\alpha=0.94$ ($z=0$) and $\alpha=0.8$ ($z=1$) as the overall best-fit values for the slopes that provide reasonable fits for all of our host halo mass bins. Holding $\alpha$ fixed for each redshift, we re-fit the amplitude $A_M$ from Equation~\ref{eq:fittingfn} to our mass function for each bin. The lower panels of Figure \ref{fig:massfnLJ} show the ratio of our core mass function to the resulting fitting function for each redshift and host halo mass bin. The best-fit parameter value for the normalization $A_M$ is displayed in each panel.

Equation \ref{eq:fittingfn}, with slope $\alpha$ fixed and the normalization $A_M$ allowed to vary, fits the data points well for both redshifts and for each mass bin for which $\log(m/M)>\log(100 m_{p,\mathrm{LJ}}/\langle M \rangle)$. This result increases our confidence in considering cores with a mass above a limit of 100 particles to be reliable models of substructure (see Section~\ref{sec:resolutiontests}). Moreover, we confirmed that fitting the core mass function for both smaller simulations (SV and HM), yielded similar results to our presented values, further demonstrating consistency between the runs.

In the literature, slope values in the range of about 0.7 to 1.1 have been found by fitting the subhalo mass function computed from N-body simulations \citep{ghigna2000,helmi02,delucia04,shaw06,diemand07,angulo09,boylan10,giocoli10,gao12}. We note that a number of factors can possibly lead to variations in the slope values, including differences in the host halo mass definition (we use an FOF, $b=0.168$ mass while some others have used $M_{200}$ or virial mass definitions), differences in the form of the power law fitting function, and differences in the definition of the subhalo mass function (e.g. cumulative and differential SHMFs). 

One important consideration when comparing our slope measurements comes from our mass-loss model parameter selection. As described in Section \ref{sec:parameterexploration}, we chose parameters that provide the best overall fit to the SHMF at redshifts $z=0$ and $z=1$. This result provided reasonable results for both redshifts. However, there was a measured variation of best-fit parameters when individually fitting each redshift independently (indicating a slight redshift-dependence that was not fully captured by our simple model). We have repeated the slope measurements at $z=0$ for the HM and SV simulations using mass-model parameter values that provide an optimal fit for our $z=0$ SHMFs.
We found the fitted slopes of some mass bins to be closer to the value given in \cite{JvdB2016a} of $\alpha=0.86$, as expected given our measured subhalo finder agreement in Section \ref{sec:SHF}.
Nevertheless, the empirical measurements we found of $\alpha=0.94$ ($z=0$) and $\alpha=0.8$ ($z=1$) using our fiducial model parameters are consistent with the range in past works, and allow us to maintain a simple self-contained model without the addition of redshift-dependent modifications. 

\section{Summary and Outlook}
\label{sec:summary}

In this paper, we introduced SMACC, Subhalo Mass-loss Analysis using Core Catalogs, a methodology for enabling the use of halo cores as an efficient substitute for subhalo catalog generation. SMACC provides mass estimates for halos after they have merged to become substructures within other halos. The mass-loss model uses the infall mass of a merging halo as a starting point, and then evolves the mass via a simple two-parameter model. The core tracking approach (summarized in Section~\ref{sec:coreconcept}) in combination with SMACC provides a very fast, scalable method to follow the evolution of substructure within halos over time. We apply the SMACC procedure to the Last Journey simulation (reported in Section~\ref{sec:results}), which is described in the first paper of the Last Journey series~\citep{LJ1}.

We established our approach in Section~\ref{sec:massmodeling}, where we fit results obtained using SMACC to the subhalo mass function measured on the same set of halos, allowing us to choose best fit parameters where the model agreed well with the subhalo results at $z=0$ and $z=1$.  (The results from our subhalo finder are consistent with those from other commonly used finders as shown in Section~\ref{sec:simulations}.) Importantly, we carried out mass resolution tests (Section~\ref{sec:resolutiontests}) to demonstrate the validity of our approach with respect to mass thresholds as well as a range of efficacy tests (Section~\ref{sec:validation}). These tests included the investigation of the robustness of SMACC towards different host halo properties, in particular the host halo concentration as discussed in Section~\ref{sec:Concentration}, finding that the simple mass-loss model holds for different host halo sub-populations based on concentration. We also carried out an individual subhalo-to-core matching comparison in Section~\ref{sec:CoreSHMatch}, showing good agreement between the measurements. Finally, in Section~\ref{sec:Cosmology} we tested the robustness of our approach against changes in the cosmological models as currently allowed by observational constraints and found essentially no cosmology dependence.

One of the primary aims for the development of SMACC is the generation of merger trees with embedded substructure information acquired through core tracking. These merger trees can then be used in many different science studies, such as input to SAMs or empirical models for the galaxy-halo connection that rely on using detailed information of the evolution of halo substructure over time. In this paper, we have focused on demonstrating that SMACC can successfully deliver an alternative approach for building detailed halo substructure models. We have taken the first step of subhalo mass modeling and have not considered the treatment of core mergers as potential proxies for galaxy mergers and of core disruption as a possible proxy for galaxy disruption.
The availability of information on the distribution of particles within a core can be applied to monitor such processes. In the specific context of modeling the galaxy distribution in clusters,~\cite{korytov} describe a core merging and disruption procedure that utilizes core particle distribution data. We will incorporate and further develop such an approach in future work to provide more simulation-based inputs to semi-analytic modeling approaches for galaxy formation.

\begin{acknowledgments}

We thank Andrew Hearin and Malin Renneby for many useful discussions, and Claude-Andr\'e Faucher-Gigu\`ere for valuable comments. We acknowledge important contributions from Hal Finkel, Adrian Pope and Tom Uram for previous work on HACC and the first paper in the Last Journey series.  Argonne National Laboratory's work was supported under the U.S. Department of Energy contract DE-AC02-06CH11357.  Awards of computer time were provided by the ASCR Leadership Computing Challenge (ALCC) program. This research used resources of the Argonne Leadership Computing Facility at the Argonne National Laboratory, which is supported by the Office
of Science of the U.S. Department of Energy under Contract
No. DE-AC02-06CH11357. We are indebted to the ALCF team for their
outstanding support and help to enable us to carry out a simulation at
this scale. This research also used resources of the Oak Ridge Leadership Computing Facility, which is a DOE Office of Science User Facility supported under Contract DE-AC05-00OR22725.

\end{acknowledgments}

\end{document}